\documentclass[prl,aps,reprint,a4paper,superscriptaddress,notitlepage,preprintnumbers,showpacs,showkeys]{revtex4-1}
\usepackage{graphicx}
\usepackage{dcolumn}
\usepackage{bm}
\usepackage{amsmath}
\usepackage{amssymb}
\usepackage{booktabs}
\usepackage{bbold}
\usepackage{color,xcolor}
\usepackage{epsfig}
\usepackage{epstopdf}
\usepackage{hyperref}
\hypersetup{pdfstartview=FitH, pdfremotestartview=FitR, colorlinks=false, linkbordercolor=red, linkcolor=green, citecolor=blue, pdfborderstyle={/S/U/W 1}, pdfnewwindow=true}
\newcommand{\msc}[1]{\text{\textsc{#1}}}
\newcommand{\bra}[1]{\langle #1|}
\newcommand{\ket}[1]{|#1\rangle}
\newcommand{\braket}[2]{\langle #1 | #2 \rangle}
\newcommand{\Ignore}[1]{}
\newcommand{\im}{{\rm Im}}
\newcommand{\re}{{\rm Re}}
\def\SI{\cite{supI}}
\def\BSGnrs{\ref{Fig1}}
\def\dplGNRsloss{\ref{Fig2}}
\def\DopedZGNR{\ref{Fig3}}
\def\DopedAGNR{\ref{Fig4}}
\def\optgeom{\ref{FigS1}}
\def\optgeomB{\ref{FigS2}}
\def\Intrinsic{\ref{FigS3}}
\def\DopedZGNRB{\ref{FigS4}}
\def\DopedAGNRB{\ref{FigS5}}
\def\DopedAGNROB{\ref{FigS6}}
\def\THzAGNRB{\ref{FigS7}}
\def\AGNRtemp{\ref{FigS8}}

\newcommand{\old}[1]{}
\newcommand{\InT}[1]{#1}
\begin{document}
\title{Plasmon Modes of Graphene Nanoribbons with Periodic Planar Arrangements}
\author{\firstname{C.}~\surname{Vacacela Gomez}}
\affiliation{Dipartimento di Fisica, Universit\`{a} della Calabria, Via P. Bucci, Cubo 30C, I-87036 Rende (CS), Italy}
\affiliation{INFN, Sezione LNF, Gruppo Collegato di Cosenza, Cubo 31C, I-87036 Rende (CS), Italy}
\author{\firstname{M.}~\surname{Pisarra}}
\affiliation{Dipartimento di Fisica, Universit\`a della Calabria, Via P. Bucci, Cubo 30C, I-87036 Rende (CS), Italy}
\affiliation{Departamento de Qu\'{\i}mica, Universidad Aut\'{o}noma de Madrid, Calle Francisco Tom\'{a}s y Valiente 7 (M\'{o}dulo 13), E-28049 Madrid, Spain}
\author{\firstname{M.}~\surname{Gravina}}
\affiliation{Dipartimento di Fisica, Universit\`a della Calabria, Via P. Bucci, Cubo 30C, I-87036 Rende (CS), Italy}
\affiliation{INFN, Sezione LNF, Gruppo Collegato di Cosenza, Cubo 31C, I-87036 Rende (CS), Italy}
\author{\firstname{J. M.}~\surname{Pitarke}}
\affiliation{CIC nanoGUNE, Tolosa Hiribidea 76, E-20018 Donostia - San Sebastian, Basque Country, Spain}
\affiliation{Materia Kondentsatuaren Fisika Saila, DIPC, and Centro Fisica Materiales CSIC-UPV/EHU, 644 Posta Kutxatila, E-48080 Bilbo, Basque Country, Spain}
\author{\firstname{A.}~\surname{Sindona}}
\email{corresponding author: antonello.sindona@fis.unical.it}
\affiliation{Dipartimento di Fisica, Universit\`{a} della Calabria, Via P. Bucci, Cubo 30C, I-87036 Rende (CS), Italy}
\affiliation{INFN, Sezione LNF, Gruppo Collegato di Cosenza, Cubo 31C, I-87036 Rende (CS), Italy}

\begin{abstract}
Among their amazing properties, graphene and related low-dimensional materials show quantized charge-density fluctuations--known as plasmons--when exposed to photons or electrons of suitable energies.
Graphene nanoribbons offer an enhanced tunability of these resonant modes, due to their geometrically controllable band gaps.
The formidable effort made over recent years in developing graphene-based technologies is however weakened by a lack of predictive modeling approaches that draw upon available {\it ab initio}  methods.
An example of such a framework is presented here, focusing  on narrow-width graphene nanoribbons organized in periodic planar arrays.
Time-dependent density-functional calculations reveal unprecedented plasmon modes of different nature at visible to infrared energies.
Specifically, semimetallic~(zigzag) nanoribbons display an intraband plasmon following  the energy-momentum dispersion of a two-dimensional electron gas.
Semiconducting~(armchair) nanoribbons are instead characterized by two distinct intraband and interband plasmons, whose fascinating interplay is extremely responsive to either injection of charge carriers or increase in electronic temperature.
These oscillations share some common trends with recent nanoinfrared  imaging of confined edge and surface plasmon modes detected in graphene nanoribbons of $100$-$500$~nm width.
\end{abstract}
\pacs{73.20.Mf,73.22.Lp,73.22.Pr}
\keywords{graphene, graphene nanoribbon, plasmonics, time-dependent density functional theory}
\maketitle
\phantomsection
\addcontentsline{toc}{section}{Introduction}
Plasmons are quantized oscillations of the valence electron density in metals, metal-dielectric interfaces and nanostructures, being usually excited by light or electron-beam radiation.
Plasmon-related technologies are expected to receive a burst from nanocarbon architectures\InT{~\cite{Lin2009,Shuba2009,bao2012graphene,sensale2012graphene,ThongrattanasiriACSNANO2012,ib00,ib01,garcia2014graphene,brun2015,fei2015},}
due to one of the fascinating features of monolayer graphene~(MG)~\cite{neto2009electronic}, i.e., its extrinsic plasmon modes at terahertz~(THz) frequencies\InT{~\cite{ju2011graphene,zhou2012atomically,yan2013damping,liou2015pi,cupolillo2015substrate,pisarra2014acoustic,PRB_Pisarra,Sind_SPRING,ibxx}.}
These show much stronger confinement, larger tunability and lower losses~\cite{christensen2011graphene} compared to conventional plasmonic materials, such as silver or gold.
Nowadays plasmons are launched, controlled, manipulated and detected in a variety of graphene-related materials and heterostructures, which suggests that graphene-based plasmonic devices are becoming closer to reality, with the potential to operate on the ``{\it THz gap}'', forbidden by either classical electronics or photonics~\cite{sensale2012graphene,GraBN_plasmon,Tong2015}.
Plasmons with widely tunable frequencies have been observed in graphene nanoribbons~(GNRs)--from the nano- to microrange in width~\cite{ju2011graphene,yan2013damping,nikitin2012surface,Feinanolett.5b03834}.
On the theoretical side, density functional and tight-binding~(TB) approaches have explored the electronic structure of zigzag and armchair GNRs, with particular attention to the band-gap values of the intrinsic systems, being a major control factor of their plasmonic properties~\cite{son2006energy,zGNRGAP2007,Dubois2009,kan2008half,tao2011spatially}.
Far fewer studies have been focused on plasmon resonances in GNRs using either a semiclassical electromagnetic picture~\cite{popov2010oblique} or a TB scheme~\cite{BreyETAl2006,andersen2012plasmon,ThongrattanasiriACSNANO2012}, and  specializing to THz frequencies.
A comprehensive characterization of the dielectric properties of such systems is, however, lacking.

Here, we provide an {\it ab initio} study of plasmon excitations in regular planar arrays of GNRs, sorting a wide range of frequencies, from the lower THz to extreme ultraviolet~(UV).
We use time-dependent~(TD) density functional theory~(DFT) in random-phase approximation~(RPA), emphasizing the 4ZGNR and 5AGNR geometries, which are respectively characterized by four zigzag chains~[Fig.~{\BSGnrs}(a)] and five dimer lines~[Fig.~{\BSGnrs}(c)] across the GNR width~\cite{son2006energy}.

\InT{
The dangling bonds of each GNR array are passivated by hydrogen atoms on both sides, with the C-C and C-H bond lengths being fixed to their nominal values, which differ by less than 1$\%$ from the corresponding geometrically optimized values~(Figs.~{\optgeom}~and~{\optgeomB} in~{\SI}).
The GNR arrays are separated by an in-plane vacuum width of $15$~{\AA\;}~[Figs.~{\BSGnrs}(a) and~{\BSGnrs}(c)], while periodic boundary conditions are used for the direction parallel to the GNR axis, which mimics a situation of long suspended ribbons with fixed edges on the far ends.}
\phantomsection
\addcontentsline{toc}{section}{GNR Electronic Structure}
The equilibrium electronic structures of the systems~(Sec.~I in~{\SI}) are computed using the local-density approximation~(LDA~\cite{perdew1981self}) with norm-conserving pseudopotentials~\cite{troullier1991efficient} and the plane-wave~(PW) basis~\cite{Abinit,monkhorst1976special}.
The three-dimensional periodicity required by PW-DFT is generated by replicating the GNR arrays over an out-of-plane distance $L$ of $15$~\AA, which ensures negligible overlap~(but not negligible interaction) of charge density between the replicas.

Accordingly, the different geometry of the assemblies~[Figs.~{\BSGnrs}(a) and~{\BSGnrs}(c)] produces electronically distinct band dispersions and densities of states~(DOS).
4ZGNR~[Fig.~{\BSGnrs}(b)] appears as a semimetal with the valence and conduction bands overlapping close to the $X$ point.
The quasiflat dispersions near the intrinsic Fermi-level $E_F$ give rise to strong peaks in the DOS, as opposite to MG where the linear dispersing valence and conduction levels yield a vanishing DOS at $E_F$.
5AGNR~[Fig.~{\BSGnrs}(d)] is a semiconductor with the valence and conduction electrons having parabolic-like dispersions around a small gap of $\sim 0.36$~eV at the $\Gamma$ point that result in two peaks in the DOS.
It should be noted that local spin density calculations suggest the opening of a band gap larger than $0.1$~eV in ZGNRs~\cite{son2006energy,zGNRGAP2007,Dubois2009}.
Additionally, GW approaches predict larger band gaps in both ZGNRs and AGNRs by roughly $1$~eV with respect to local density calculations~\cite{zGNRGAP2007}.
Nonetheless, band-gap values of the same order of the LDA band gap of 5AGNR have been measured for some GNRs as wide as about $20$~nm grown on Au(111)~\cite{tao2011spatially}.
Thus, the application of an RPA scheme to the LDA band structure of 5AGNR can be of help in interpreting plasmon measurements on currently synthesized GNR structures~\cite{Feinanolett.5b03834}.
Complementarily, the LDA analysis of a virtually gapless GNR, i.e., 4ZGNR, is particularly instructive to emphasizing the different role played by doping.
\begin{figure}[h]
\centerline{\includegraphics[width=0.48\textwidth]{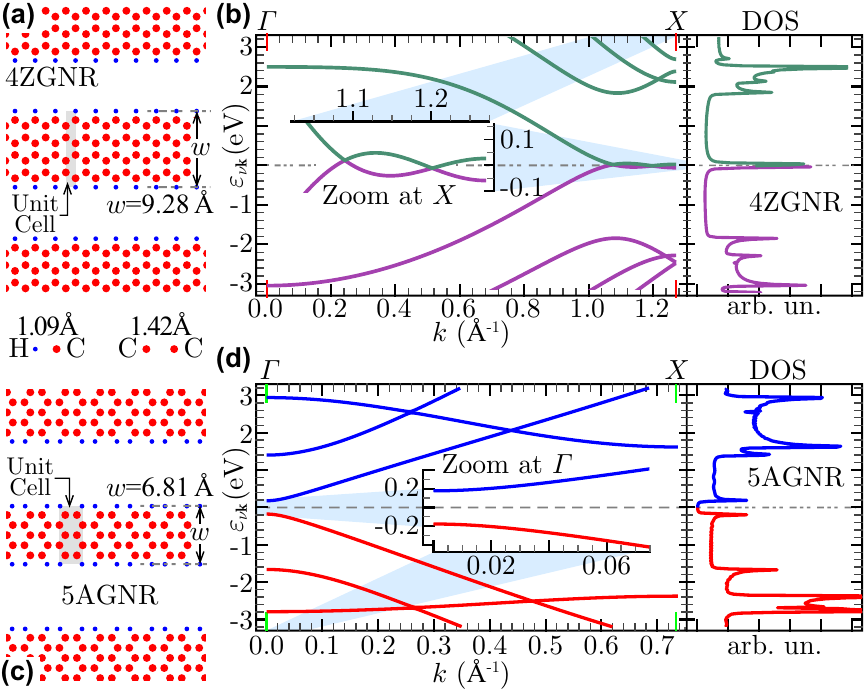}}
\vskip -10pt
\caption{
Geometry~[(a) and (c)], band energies~(zoomed at $E_F=0$) and DOS [(b) and (d)] for 4ZGNR [(a) adn (b)] and 5AGNR [(c) and (d)].
\label{Fig1}}
\end{figure}

\phantomsection
\addcontentsline{toc}{section}{TDDFT+RPA Approach}
The starting point of our TDDFT approach is the density-density response function of {\it noninteracting} electrons in the GNRs, as given by the Adler-Wiser formula~\cite{adler1962quantum,wiser1963dielectric}
\begin{align}
\label{AdlWi}
&\chi _{{\bf G} {\bf G}'}^{0} =
\frac{2}{\Omega}
\sum_{{\bf k},\nu,\nu'}
\frac{
(f_{\nu {\bf k}}-f_{\nu' {\bf k}+{\bf q}})
\rho_{\nu\nu'}^{{\bf k} {\bf q}}({\bf G})\,
\rho_{\nu\nu'}^{{\bf k} {\bf q}}({\bf G}')^{\ast}
}{
\omega+\varepsilon _{\nu {\bf k}}-\varepsilon_{\nu'{\bf k}+{\bf q}}+{\rm i}\eta },
\end{align}
which is a corollary of the Kubo formula for a periodic system with associated reciprocal lattice and Brilloiun Zone~(BZ)~\cite{kubo1957statistical}.
Hartree atomic units are used throughout this work, unless otherwise stated.

In Eq.~\eqref{AdlWi} the electron energies $\varepsilon_{\nu {\bf k}}$ and states $\ket{\nu {\bf k}}$ are indexed by the band number $\nu$ and the wave vector ${\bf k}$ 
in the first BZ.
These are taken to be the KS-eigensystems of our PW-DFT approach, leading to the electronic structure of Figs.~{\BSGnrs}(b) and~{\BSGnrs}(d).
The KS wave functions, normalized to unity in the crystal volume ${\Omega}$, are expressed as linear combinations of PWs that depend on the reciprocal-lattice vectors ${\bf G}$ associated to the replicated GNR lattices~(Sec.~I~and~II in~{\SI}).
The correlation matrix elements are given by
$\rho_{\nu\nu'}^{{\bf k}{\bf q}}({\bf G})=
\bra{\nu {\bf k}}
e^{-{\rm i}({\bf q}+{\bf G})\cdot\mathbf{r}}
\ket{\nu'{\bf k}+{\bf q}}$.
The population of single-particle levels is established by the Fermi-Dirac distribution $f_{\nu \mathbf{k}}$, which we evaluate by sampling temperatures from $300$ to $900$~K.
The factor of $2$ accounts for the spin degeneracy, while $\eta$ is a small~(positive) lifetime broadening parameter~\cite{PhysRevB.68.205106}.

Polarization effects are activated by a test electron or photon with incident energy $\omega$ and in-plane momentum ${\bf q}$ that weakly perturbs our systems.
These 
are described by the density-density response function of {\it interacting} electrons, which can
be obtained in the TDDFT framework as
$
\chi_{{\bf G}{\bf G}'}= \chi^0_{{\bf G}{\bf G}'} + (\chi^0 v \chi)_{{\bf G}{\bf G}'}\label{chi}
$~\cite{TDDFT,onida2002electronic}.

In the RPA, one neglects short-range exchange-correlation effects by simply replacing the {\it unknown}
$v$ by the bare Coulomb terms $v^0_{{\bf G}{\bf G}'}=4\pi\delta_{{\bf G}{\bf G}'}/|{\bf q} + {\bf G}|^2$.
A serious drawback stems from the long-range character of the Coulomb potential, which allows nonnegligible interactions between repeated planar arrays even at large distances.
To cut off this unwanted phenomenon, we replace $v^0_{{\bf G}{\bf G}'}$ by the truncated Fourier integral~\cite{despoja2012ab,despoja2013two,despoja2015three,PRB_Pisarra,Sind_SPRING}
\begin{equation}
\label{v2dC}
v_{{\bf G}{\bf G}'} = \int_{-L/2}^{L/2} dz \int_{-L/2}^{L/2} dz' e^{{\rm i} G z} \bar{v}^0_{{\bf g} {\bf g}'}(z,z') e^{-{\rm i} G' z'},
\end{equation}
where $\bar{v}_{{\bf g} {\bf g}'}^0$ is the Fourier transform of
$v^0_{{\bf G}{\bf G}'}$ along the out-of-plane axis, while ${\bf g}$ and $G$ denote the in-plane and out-of-plane components of ${\bf G}$.

Within linear response theory, the inelastic cross section corresponding to a process where the external perturbation creates an excitation of energy $\omega$ and wave vector ${\bf q}+{\bf G}$ is related to the diagonal elements of the inverse dielectric matrix:
$
(\epsilon^{-1})_{{\bf G} {\bf G}'}=\delta_{{\bf G}{\bf G}'}+(v \chi)_{{\bf G} {\bf G}'}.
\label{epsm1}
$
Collective excitations~(plasmons) are dictated by the zeros in the real part of the macroscopic dielectric function~(permittivity):
$
\epsilon^M=1/(\epsilon^{-1})_{\mathbf{0}\mathbf{0}}.
$
The so-called energy-loss~(EL) function is proportional to the imaginary part of the inverse permittivity:
$
E_{\msc{loss}}=-\im[(\epsilon^{-1})_{\mathbf{0}\mathbf{0}}]$.
Nonlocal field effects are included in $E_{\msc{loss}}$ through the off-diagonal elements of $\chi_{{\bf G}{\bf G}'}$~\cite{PhysRevLett.100.196803}.

\phantomsection
\addcontentsline{toc}{section}{Intrinsic GNRs}
The peculiar electronic structure of 4ZGNR~[Fig.~{\BSGnrs}(b)] and 5AGNR~[Fig.~{\BSGnrs}(d)], as compared to the well-known band dispersion of MG, is reflected in the EL spectra of the intrinsic systems shown in Fig.~{\dplGNRsloss}.
Undoped 4ZGNR and 5AGNR have two high-energy excitations~(for $\omega>3$~eV) that follow one-electron transitions connecting the ${\bf k}$ points with high DOS in the $\pi$-$\pi^*$, $\sigma$-$\pi^*$ and $\pi$-$\sigma^*$ bands.
These are counterparts to the $\pi$ and $\sigma$-$\pi$ {\it interband} plasmons observed in intrinsic MG~[Figs.~{\dplGNRsloss}(a) and~{\dplGNRsloss}(b)], few-layer graphene and graphite~\cite{PhysRevB.77.233406,PhysRevB.88.075433,PhysRevB.90.235434,taft}.
Specifically, the $\pi$ and $\pi$-$\sigma$ structures of the GNRs exhibit a discontinuous dispersion vs ${\bf q}$ and $\omega$, as they are split into more branches~[Figs.~{\dplGNRsloss}(c) and {\dplGNRsloss}(d), plus Figs.~{\Intrinsic}(c) and {\Intrinsic}(d) in~\SI].
This is due to the finite width of the GNRs in the periodic array, generating several one-dimensional bands of $\pi$ and $\sigma$ character.
\begin{figure}[h]
\centerline{\includegraphics[width=0.49\textwidth]{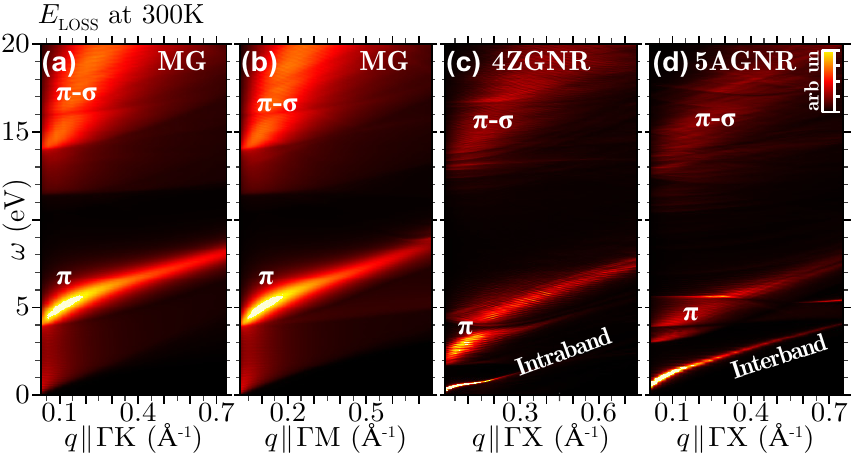}}
\vskip -10pt
\caption{$E_{\msc{loss}}$ vs $\omega<20$~eV and $q<0.8$~{\AA}$^{-1}$ for intrinsic MG (a,b), 4ZGNR (c) and 5AGNR (d).
\label{Fig2}}
\end{figure}

The number and dispersion of these bands is also strongly influenced by the GNR width, chirality and in-plane distance: the wider the ribbon the more regular the high-energy interband peaks, which approach the $\pi$ and $\sigma$-$\pi$ peaks of MG as the ribbon width tends to infinity~[Figs.~{\dplGNRsloss}(a) and {\dplGNRsloss}(b), plus Figs.~{\Intrinsic}(a) and {\Intrinsic}(b) in~\SI].
Then, the main designing ``\emph{ingredients}'' of the GNR arrays may be finely tuned to reach a specific energy for the $\pi$ and $\sigma$-$\pi$ excitations, which in turn may be used to change the response of a GNR based device working in the visible~(VIS) to UV regime.
The low-energy ends of the spectra~(for $\omega < 3$~eV) exhibit an extra peak in both metallic and semiconducting GNRs, which is strictly absent in MG at the absolute zero.
These structures are more detailed in Fig.~{\DopedZGNR}(a), {\DopedZGNR}(e), {\DopedAGNR}(a) and {\DopedAGNR}(e) below.
The large DOS value close to $E_{F}$ in 4ZGNR~[Fig.~{\BSGnrs}(b)] yields a concentration of $n^*=3.96{\times}10^{12}$~cm$^{-2}$ conduction electrons, which allows the appearance of an {\it intraband} plasmon where the charge carriers located on each ribbon of the array oscillate as a single 2D gas~[Figs.~{\dplGNRsloss}(c), {\DopedZGNR}(a) and Figs.~{\Intrinsic}(c), {\DopedZGNRB}(a) in~\SI].
This observation is confirmed by the typical square-root-like dispersion of two-dimensional plasmons~\cite{RevModPhys.54.437} for low $q$~[Fig.~{\DopedZGNRB}(d) in~\SI] that has been mostly observed in extrinsic MG~\cite{pisarra2014acoustic}, which even in the intrinsic case allows for a weak intraband mode at room temperature assisted by the conduction-electron concentration $n^*=1.15{\times}10^{11}$~cm$^{-2}$.

On the other hand, the energy gap at $E_{F}$ in 5AGNR~[Fig.~{\BSGnrs}(d)] yields a negligibly small intraband mode due to the tiny concentration of conduction electrons at room temperature~($n^*=8.70{\times}10^{8}$~cm$^{-2}$).
The latter is not detectable in Figs.~{\dplGNRsloss}(d) and {\DopedAGNR}(a), but can be, in principle, characterized~(Sec.~III in~\SI).
In contrast, an interband two-dimensional plasmon is clearly recorded in the low-energy spectrum of 5AGNR, as testified by the intense signal in Figs.~{\dplGNRsloss}(d), {\DopedAGNR}(a) and ~{\DopedAGNR}(d); this corresponds to a collective mode that is triggered by transitions between the valence and conduction DOS peaks at $\Gamma$~[Fig.~{\BSGnrs}(d)].
The collective nature of the {\em newly} detected modes in 4ZGNR and 5AGNR is proved in
Figs.~{\DopedZGNR}(e) and~{\DopedAGNR}(e), respectively, where we see that each excitation peak in the EL spectrum corresponds to a zero in the real-permittivity
$\re(\epsilon^{\msc{m}})$, at a frequency where the imaginary-permittivity $\im(\epsilon^{\msc{m}})$ is small.
\Ignore{These excitations are indeed genuine plasmons, which can be used in practical THz applications.}
Moreover, as is the case for the high-energy $\pi$ excitations, these low-energy modes arise from transitions involving the $\pi$-bands, which means that their intensities and energy-momentum dispersions can be modulated according to external factors that change the band levels, such as the already mentioned ribbon width, in-plane distance and chirality.
\begin{figure}[h]
\centerline{\includegraphics[width=0.49\textwidth]{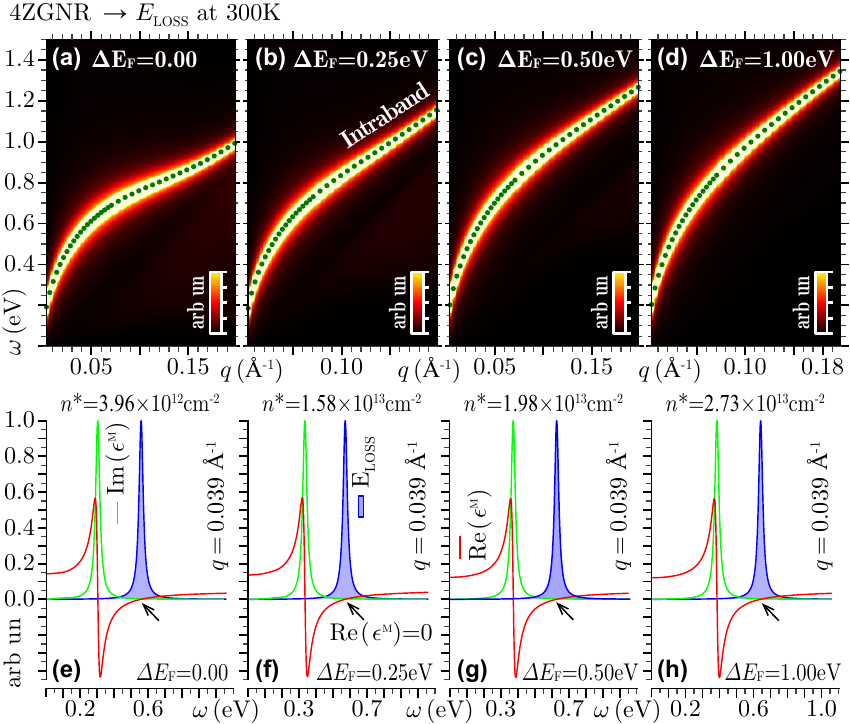}}
\vskip -10pt
\caption{EL spectrum and complex permittivity of intrinsic and extrinsic 4ZGNRs at room temperature.
(a-d) $E_{\msc{loss}}$ vs $\omega<1.5$~eV and $q < 0.11$~{\AA}$^{-1}$.
(e-h) $\re(\epsilon^{\msc{m}})$, $\im(\epsilon^{\msc{m}})$ and $E_{\msc{loss}}$ vs $\omega<1.5$~eV at $q=0.039$~{\AA}$^{-1}$.
In (a-d) the same color code or intensity scale as in Fig.~{\dplGNRsloss} is used, with the green dots denoting the $(\omega,q)$ dispersion of the intraband plasmon.
\label{Fig3}}
\end{figure}

Let us now see how the dielectric properties of the GNR arrays behave with injecting or ejecting electrons by doping or gating. Extrinsic systems are simulated here by slightly changing the level populations in
Eq.~\eqref{AdlWi}, in such a way that band dispersions and single-particle KS orbitals are negligibly altered by the applied variations of the $f_{\nu{\bf k}}$ factors.
For doping levels $\Delta E_F$ not larger than $\sim 1$~eV the high-energy end~($\omega>3$~eV) of our EL spectra is practically unaffected.
On the contrary, unprecedented new features are recorded at the low-energy end~($\omega<3$~eV).

\InT{In Fig.~{\DopedZGNR}(a)-{\DopedZGNR}(d)~(and Fig.~{\DopedZGNRB}(a)-{\DopedZGNRB}(c) in~\SI) we show the low-$\omega$ and low-${\bf q}$ region of the EL spectrum of 4ZGNR arrays, zooming on the undoped case and analyzing three positive doping levels, whose charge-carrier concentrations are consistent with the measurements of  Ref.~\cite{ib03}.}
We observe a single dispersive structure, the intraband plasmon, which is a genuine collective mode, with the EL peak corresponding to a zero in $\re(\epsilon^{\msc{m}})$ and a small value of $\im(\epsilon^{\msc{m}})$~[Fig.~{\DopedZGNR}(e-h)].
We also notice minor differences in the four EL spectra, with the plasmon energy slightly increasing with increasing $\Delta E_F$ up to 1.0~eV~[Fig.~{\DopedZGNRB}(d) in~\SI].

\phantomsection
\addcontentsline{toc}{section}{Doped GNRs}
More interesting features are observed in doped 5AGNR arrays, whose low-$\omega$ and low-${\bf q}$ response is shown in Fig.~{\DopedAGNR}.
In the undoped case, a single dispersive peak is detected that represents an interband plasmon, which follows coherent one-electron transitions between valence and conduction states~[Figs.~{\DopedAGNR}(a), {\DopedAGNR}(d) and Fig.~{\DopedAGNRB}(c) in~\SI].
When a small doping is introduced~($\Delta E_F=-0.2,0.3$~eV) the conduction-electron or valence-hole concentration bursts from $\pm 10^9$ to $\pm 10^{12}$~cm$^{-2}${, a value reported in previous experiments~\cite{ib02}.} Then, another dispersive peak appears due to a clearly resolved intraband plasmon~[Figs.~{\DopedAGNR}(b), {\DopedAGNR}(c), {\DopedAGNR}(f), {\DopedAGNR}(g) and Figs.~{\DopedAGNRB}(b), {\DopedAGNRB}(e) in~\SI].

For these low wave vectors~($q<0.02$~{\AA}$^{-1}$), the intraband mode is the most intense contribution, while the interband plasmon is depressed because the doping partially fills the conduction band near $\Gamma$ thus inhibiting quasivertical $(q \to 0,\omega)$ interband transitions.
In the $0.02<{q}<0.06$~{\AA}$^{-1}$ region, both the intraband and interband plasmons coexist.
At larger values of $q$, the interband plasmon becomes the most intense peak while the intraband plasmon is strongly damped.

A slightly larger value of the doping~($\Delta E_F=0.4$~eV) leads to an even more intriguing situation: the single dispersive peak visible in Figs.~{\DopedAGNR}(d) and~{\DopedAGNR}(h) has a {\bf double nature}, as testified by the kink in peak dispersion and the abrupt decrease in intensity~(increase in width) found at $q\sim0.05$~{\AA}$^{-1}$~[Fig.~{\DopedAGNRB}(f) in~\SI].
Indeed, interband transitions between the high-DOS points of Fig.~{\BSGnrs}(d) for $q<0.04$~{\AA}$^{-1}$ are strongly quenched by electron population of conduction levels; thus, the intense peak showing the $\sqrt{q}$ dispersion is mostly originated by the intraband plasmon.
Conversely, for $q>0.04$~{\AA}$^{-1}$ the intraband plasmon enters a region where it is damped by interband transitions; as a result, most of the spectral weight is concentrated on the interband plasmon, while the overdamped intraband plasmon only appears as a faint peak.
\begin{figure}[h]
\centerline{\includegraphics[width=0.49\textwidth]{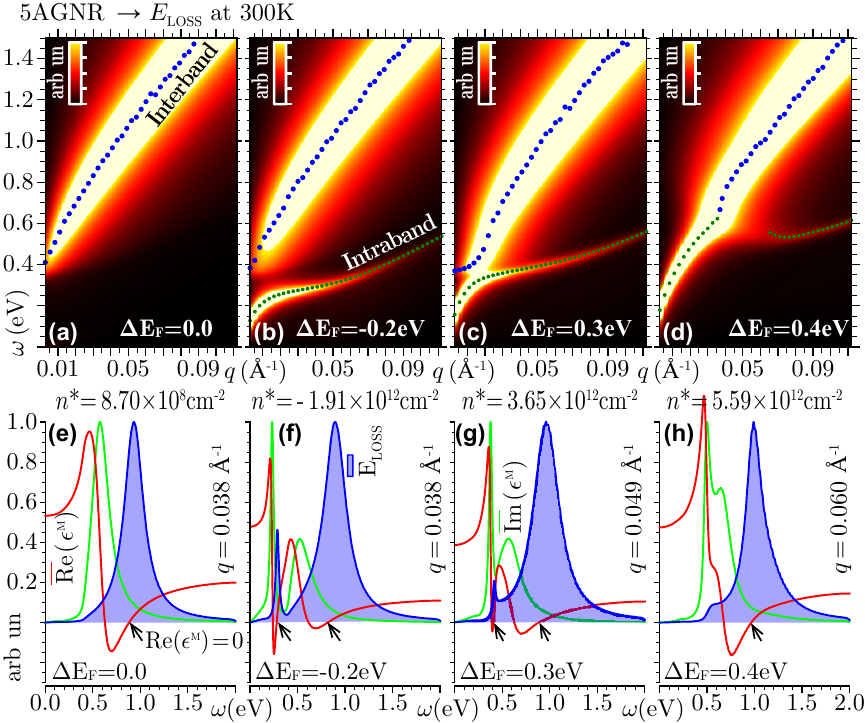}}
\vskip -10pt
\caption{Dielectric response and plasmon dispersions of intrinsic and extrinsic 5AGNRs at $T=300$~K. (a-d) $E_{\msc{loss}}$ vs $\omega<1.5$~eV and $q < 0.11$~{\AA}$^{-1}$ with the same intensity scale as Fig.~{\dplGNRsloss} and Fig.~{\DopedZGNR}(a)-{\DopedZGNR}(d). The blue and green dots mark the interband  and intraband plasmon $(\omega,q)$ dispersions, respectively.
\label{Fig4}}
\end{figure}
Another remarkable effect is the high sensitivity of the intraband plasmon to the type of doping; opposite doping levels, such as the $\Delta E_F=\pm 0.2$ and $\Delta E_F=\pm 0.3$ values of Fig.~{\DopedAGNRB}(h) in~\SI, produce significantly different charge-carrier concentrations and plasmon dispersion curves for energies larger than $0.1$~eV and transferred momenta above $0.02$~{\AA}$^{-1}$.
Such a sensitivity is ascribed to the slight asymmetry of the valence and conduction electron levels of 5AGNR close to the band gap~[inset in Fig.~{\BSGnrs}(d)].

On the other hand, the interband plasmon is much less influenced by the doping type, because the valence and conduction DOS peaks have similar intensities~[Fig.~{\DopedAGNRB}(h) in~\SI].
The interplay of the two modes is also modulated by changes in the incident momentum direction, relative to the ribbon  axis\InT{, due to the tensor character of the GNR dielectric response}~(Fig.~{\DopedAGNROB} in~\SI).
In the lower THz region, interband transitions are forbidden by the band gap. The intraband plasmon follows an $\omega$-vs-$q$ dispersion that is consistent with the semiphenomenological relation of Ref.~\cite{popov2010oblique}, and appears at the same scale as the low-$q$ dispersions of Ref.~\cite{andersen2012plasmon}, computed from the TB approximation where 5AGNR is virtually gapless~[Fig.~{\THzAGNRB}(a) in {\SI}].
Then, the TDDFT response of narrow-width GNRs has the correct $q{\to}0$, $\omega{\to}0$ limiting behavior predicted by (non-{\it ab initio}) approaches on larger GNR structures, currently available for experiments.

As a final remark we observe that in semiconducting GNRs, due to their small band gap, the temperature plays a role in dictating the populations of the levels close to $E_F$~(Sec.~IV in~\SI).
Accordingly, the intraband plasmon mode can be triggered by working at temperatures larger than $\sim 500$~K, which may have a crucial role in relation with power consumption of nanodevices.
Charge-carrier concentrations generated by temperature increase are nevertheless much smaller than those obtained with doping or gating. For this reason, no particular interference is recorded between intraband and interband plasmon modes~(Fig.~\AGNRtemp in~\SI).

\Ignore{In summary, we have discussed the dielectric properties and plasmon dispersions in 2D arrays of GNRs scrutinizing the excitation energy regime going from the THz to the UV scale.
The use of an {\it ab initio}  strategy based on TDDFT in the RPA has let us characterize both the intrinsic and extrinsic plasmon structures of the materials, which could not have possibly been predicted by less sophisticated TB approaches. Indeed, the latter are mostly based on a two-band model for the valence and conduction states, with a semiphenomenological band gap, and do not reliably take into account of the correlation coefficients in Eq.~{\eqref{AdlWi}}.}
In summary, we have discussed the dielectric properties and plasmon dispersions in planar GNR arrays scrutinizing the excitation energy regime going from the THz to the UV scale, by an {\it ab initio}  strategy based on TDDFT+RPA.
\Ignore{At VIS to UV frequencies, we have found the two standard interband excitations of carbon-based materials, namely the $\pi$ and $\sigma$-$\pi$ plasmons, with the $\pi$ plasmon being strongly influenced by the GNR geometry.}
\phantomsection
\addcontentsline{toc}{section}{Conclusions}

On the THz regime, we have detected {\it new} collective modes of different nature.
Semimetallic GNRs display an intraband 2D plasmon with large intensity relative to the high-energy plasmons even in the intrinsic case.
Semiconducting GNRs experience a fascinating interplay of intraband and interband collective modes, whose relative intensities and dispersions are strongly influenced by the actual occupation of single-particle levels near the Fermi energy.
\Ignore{This strong sensitivity allows for a high tunability and control of the new plasmons.}

\InT{Some recent calculations have reported the existence of two extrinsic plasmons in MG and bilayer graphene~(BLG)~\cite{pisarra2014acoustic,PRB_Pisarra,Sind_SPRING,ibxx}. In particular, the plasmon coupling in BLG~\cite{ibxx} shares some common features with 5AGNR at similar doping conditions; i.e., one of the two modes disperses like $q^{1/2}$ and the other is quasivertical.}

\InT{Indeed, a first experimental evidence of an {\it edge} (interband) plasmon superimposed to a {\it conventional} (intraband) plasmon has been given} in patterned GNRs grown on Al$_2$O$_3$~\cite{Feinanolett.5b03834}. The two modes are well resolved in space on GNR samples of $480$~nm width, at a working frequency of $\sim 0.15$~eV and a doping level of $\sim 0.3$~eV.
In our narrow-width GNR, the interband and intraband features are resolved in momentum space, only.

Our calculations demonstrate that it is possible to construct new materials with plasmonic resonances that are tunable to suit a specific demand in both the VIS-UV and THz regimes, by means of chemical doping, electronic gating, and also a careful choice of the geometry.
These findings if confirmed by further experiments will widen the perspectives on applications of GNR arrays for the engineering of nanophotonic and nanoelectronic devices.
\begin{acknowledgements}
C.V.G. acknowledges the financial support of ``{\em Secretaria de Educaci\'on Superior, Ciencia, Tecnolog\'ia e Innovaci\'on}'' (SENESCYT-ECUADOR).\\
All authors thank Dr V.M.~Silkin from the University of the Basque Country for his invaluable support in developing the TDDFT code.
\end{acknowledgements}

\clearpage
\setcounter{equation}{0}
\renewcommand{\theequation}{S\arabic{equation}}
\setcounter{figure}{0}
\renewcommand{\thefigure}{S\arabic{figure}}
\setcounter{page}{0}
\renewcommand{\thepage}{\roman{page}}
\begin{widetext}
\phantomsection
\addcontentsline{toc}{section}{Supplemental Material}
\hskip 176.5pt{\bf \large Supplemental Material}

\vskip 12pt
The main text has been concerned with the dielectric response of two narrow-size graphene nanoribbons~(GNRs),  organized in periodic planar arrays with the zigzag~(Z) and armchair~(A) geometries.
Time-dependent density-functional theory has been carried out for both intrinsic~(undoped) and extrinsic~(doped, gated) systems within the random phase approximation~(RPA).
The energy loss spectra of the GNR arrays have been reported and discussed for a range of incident energies below $20$~eV and transferred momenta smaller than $0.8$~{\AA}$^{-1}$.
These supplementary notes are organized as follows:
Sec.~I provides the details on ground state computations and geometry optimizations for the GNR arrays, denoted in the main text as 4ZGNR and 5AGNR~\cite{son2006energy};
Sec.~II deals with the calculations of the plasmon structure of the intrinsic systems in comparison with that of pristine graphene;
Sec.~III gives some additional arguments on the effect of doping in 4ZGNR and 5AGNR, focusing on the energy-momentum dispersion of the intraband and interband plasmons; the semiclassical limit of the approach is validated by zooming on the lower THz region;
Sec.~IV is devoted specifically to the role played by increasing the electronic temperature on 5AGNR.
\end{widetext}

\section{Local Density Calculations and Geometry Optimization}
Density functional calculations were performed using the plane-wave~(PW) basis set~\cite{Abinit}, represented by the space functions
$
{\rm PW}_{{\bf k}+{\bf G}}({\bf r})=\Omega_0^{-1/2}e^{ i  ({\bf k}+{\bf G})\cdot {\bf r}}, \label{PW}
$
where ${\bf k}$ is a wave vector  in the first Brillouin Zone~(BZ),  ${\bf G}$ a reciprocal vector, and  ${\Omega_0}$ denotes the unit-cell volume of the real-space lattice.
The number of PWs was limited by the cut off condition:
$|{\bf k}+{\bf G}|^2/2< 25~{\rm Hartree}$.

The local density approximation~(LDA) was used, as parametrized by the Perdew-Zunger form of the uniform-gas correlation energy~\cite{perdew1981self}.
Norm-conserving pseudopotentials of the Troullier-Martins type~\cite{troullier1991efficient} were adopted.
\begin{figure}[!h]
\centerline{\includegraphics[width=0.46\textwidth]{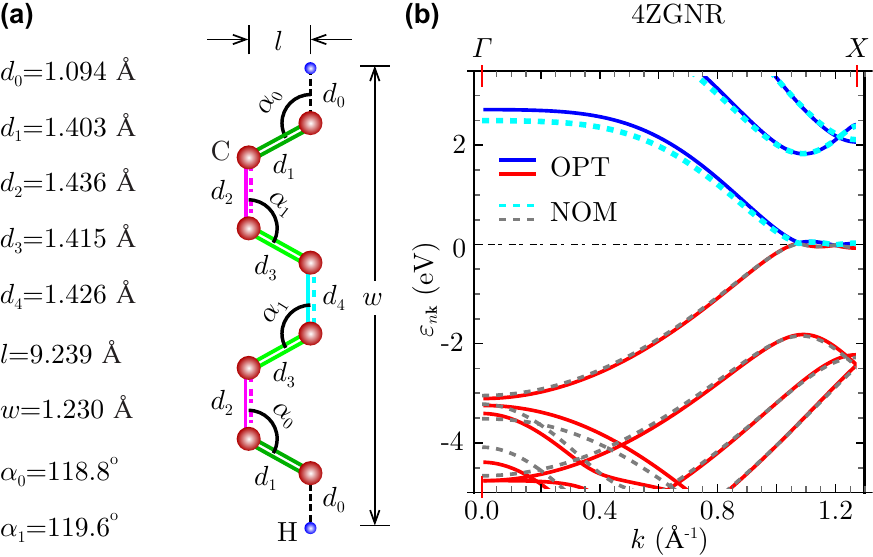}}
\vskip -10pt
\caption{
Optimized unit cell (a) and LDA electronic structure (b) for 4ZGNR.
The different bond lengths and angles are denoted $d_0,...,d_4$ and $\alpha_0,\alpha_1$,  respectively; $w$ labels the cell-width and $l$ the cell-length; the C-H bond length approximates to $1.09$~{\AA}; the average C-C bond length equals to $1.419$~{\AA}.
The band levels from the optimized~(OPT) unit cells of (a) are compared in (b) with those resulting from the nominal~(NOM) cells of Fig.~{\BSGnrs}(a), which are plotted in Fig.~{\BSGnrs}(b) of the main text.
\label{FigS1}}
\end{figure}

The basic parameters of the unit cells of 4ZGNR and 5AGNR, i.e., the C-C and C-H bond lengths and angles, were determined by geometry optimization using the Broyden-Fletcher-Goldfard-Shanno method~\cite{Abinit}.
The resulting average C-C and C-H distances turned out to be slightly different from the nominal values of  $1.42$ and $1.09$~{\AA}, respectively, used in the main text to produce the non optimized structures of Figs.~{\BSGnrs}(a) and {\BSGnrs}(c).

Periodic planar arrays of 4ZGNR and 5AGNR were then generated with both the optimized and non-optimized unit cells assuming an in-plane vacuum distance of $15$~{\AA}.
The three-dimensional periodicity required by PW-DFT was set by replicating the GNR arrays with an out of plane lattice constant of $15$~{\AA}, corresponding to the unit-cell volumes $\Omega_0 \sim 950$~{\AA}$^3$ for 4ZGNR and $\Omega_0 \sim 1470$~{\AA}$^3$ for 5AGNR.
\begin{figure}[!h]
\centerline{\includegraphics[width=0.46\textwidth]{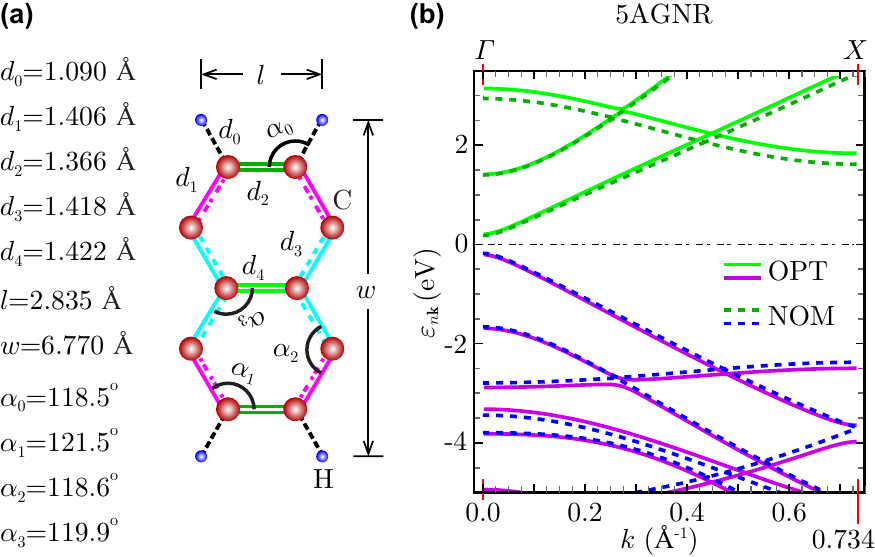}}
\vskip -10pt
\caption{
Optimized unit cells (a) and LDA electronic structure (b) 5AGNR.
The different bond lengths and angles are denoted $d_0,...,d_4$ and $\alpha_0,...,\alpha_3$,  respectively; $w$ labels the cell-width and $l$ the cell-length; the C-H bond length approximates to $1.09$~{\AA}; the average C-C bond length equals to $1.404$~{\AA}.
The band levels from the optimized~(OPT) unit cells of (a) are compared in (b) with those resulting from the nominal~(NOM) cells of Fig.~{\BSGnrs}(c), which are plotted in Fig.~{\BSGnrs}(d) of the main text.
\label{FigS2}}
\end{figure}

Both geometry optimization and ground state calculations were carried out using an unshifted~($\Gamma$-centered) Monkhorst-Pack~(MP) grid~\cite{monkhorst1976special} made of $60\times10\times1$ wave vectors, which results in a uniform sampling of the irreducible area of the first BZ~(see Fig.~{\DopedAGNROB}).

The band energies obtained with the optimized cells~[denoted OPT in Figs.~{\optgeom}(b) and~{\optgeomB}(b)] were found to differ negligibly from those of the non optimized structures~[denoted NOM in Figs.~{\optgeom}(b) and~{\optgeomB}(b)] that are also reported in Figs.~{\BSGnrs}(b) and~{\BSGnrs}(d) of the main text.

Regular planar arrays made of wider ribbons were tested as well~(though not reported here) for more systematic future studies.

Electronic structure calculations were also run on monolayer graphene~(MG) for comparison purposes~(see Figs.~{\dplGNRsloss} and~{\Intrinsic}).
In this case, the honeycomb lattice structure of C atoms~(separated by a nearest neighbor distance of $1.42$~\AA) was replicated with an out-of-plane vacuum distance of $15$~{\AA}, i.e., a unit-cell volume $\Omega_0 \sim 110$~\AA$^3$.
The first BZ of the two-dimensional material was represented on an MP grid of $60{\times}60{\times}1$ ${\bf k}$ points.

\section{Intrinsic Systems}
Energy Loss~(EL) calculations in the main text were based on the ``{\it nominal}'' cells reported in Figs.~{\BSGnrs}(a) and {\BSGnrs}(c) that have a more balanced space symmetry group with respect to the optimized cells of Figs.~{\optgeom}(a) and~{\optgeomB}(b).
With this choice, the computational burden required to sample the KS structure of the systems
is reduced to the analysis of the irreducible part of the first  BZ, ended by the $\Gamma$ and $X$ points along the GNR axis~[and shown in Figs.~{\BSGnrs}(b), {\BSGnrs}(d), {\optgeom}(b) and {\optgeomB}(b)].

In the $\Gamma{X}$ domain, we calculated the KS energies $\varepsilon_{\nu {\bf k}}$  and wave functions
\begin{equation}
\braket{{\bf r}} {\nu {\bf k}} =
N^{-1/2}\sum_{\bf G} c_{\nu {\bf k}+{\bf G}} {\rm PW}_{{\bf k}+{\bf G}}({\bf r}),
\label{wfnk}
\end{equation}
being normalized to unity in the crystal volume $\Omega=N\Omega_0$, with $N$ expressing the number of sampled {\bf k} points~(in the MP mesh) and the ${\bf G}$-sum being limited by the cut off condition set forth above.

The full plasmon structure of the intrinsic GNRs was explored at frequencies $\omega \leq 20$~eV~[Figs.~{\dplGNRsloss}(c), {\dplGNRsloss}(d), {\Intrinsic}(c) and {\Intrinsic}(d)].
Momentum values $q$ parallel to the GNR axis were considered in the range of ${\sim}0.02$ to ${\sim}0.8$~{\AA}$^{-1}$.

The converged ground state density
\begin{equation}
n({\bf r})=\sum_{\nu {\bf k}}^{{\rm occ}} |\braket{{\bf r}} {\nu {\bf k}}|^{2}
\end{equation}
was determined with $\sim 15000$ PW coefficients $\{c_{\nu {\bf k}+{\bf G}}\}$  per occupied wave function, leading to the electronic structure of Figs.~{\BSGnrs}(b) and~{\BSGnrs}(d).

$n({\bf r})$ was subsequently used in a non self-consistent run to determine the KS energies and wave functions on an MP mesh of $180\times1\times1$ ${\bf k}$ points.
$120$ bands~(indexed by $\nu$) were included to have a reliable dielectric response in the considered frequency and momentum ranges.

EL calculations for MG were run on an MP grid of $180{\times}180{\times}1$ wave vectors in the irreducible first  BZ of the two-dimensional material, including $80$ bands and $\sim 5000$ PW coefficients per wave function~[Figs.~{\dplGNRsloss}(a), {\dplGNRsloss}(b), {\Intrinsic}(a) and {\Intrinsic}(b)].

The non interacting density-density response~(or unperturbed susceptibility) $\chi^0_{{\bf G}{\bf G}'}$ was computed for MG, 4ZGNR, and 5AGNR at room temperature.
As reported in Eq.~\eqref{AdlWi} of the main text, $\chi^0_{{\bf G}{\bf G}'}$ includes all possible one-electron processes between occupied and empty band levels, separated in energy by $\varepsilon_{\nu {\bf k}}-\varepsilon_{\nu' {\bf k}+{\bf q}}$.
The dynamical screening features are provided by the retarded Green's functions $(\omega+\varepsilon _{\nu {\bf k}}-\varepsilon_{\nu'{\bf k}+{\bf q}}+ i \eta)^{-1}$, which contains the positive infinitesimal $\eta$.
Transition rates between the band levels are made of products of correlation matrix elements
\begin{align}
\rho_{\nu\nu'}^{k{\bf q}}({\bf G})=&\int_{\Omega} d^3r \braket{\nu {\bf k}}{{\bf r}}
e^{- i ({\bf q}+{\bf G})\cdot\mathbf{r}}
\braket{{\bf r}}{\nu'{\bf k}+{\bf q}}\notag \\
=&\sum_{\bf G'} c_{\nu {\bf k}-{\bf G}+{\bf G'}} c_{\nu' {\bf k}+{\bf q}+{\bf G}'},
\label{rhocc}
\end{align}
weighted by the Fermi-Dirac factors $f_{\nu{\bf k}}-f_{\nu {\bf k}+{\bf q}}$.

The unperturbed susceptibility of 4ZGNR and 5AGNR was represented with $121$ reciprocal lattice vectors, sorted in length and cut off by the condition
$
|{\bf G}|=\sqrt{|{\bf g}|^2+G^2} < 7.2{\;}{\rm\AA}^{-1},
$
with ${\bf g}$ and $G$ denoting the in-plane and out-of-plane components of ${\bf G}$, respectively.
$60$ $q$-values were sampled along  $\Gamma{X}$, and $3001$ frequencies below $20$~eV.
$\eta$ was replaced with a (finite) lifetime broadening parameter of $0.02$~eV.

The unperturbed susceptibility of MG was represented with $50$ reciprocal lattice vectors, being such that $|{\bf G}|<3.2$~{\AA}$^{-1}$~\cite{pisarra2014acoustic}. The $\Gamma{K}$ and $\Gamma{M}$ high-symmetry paths  were sampled with $60$ and $90$ $q$-values, respectively.
$3001$ $\omega$-values below $30$~eV were selected with $\eta =0.02$~eV.

Accordingly, unperturbed susceptibility matrices were calculated for all sampled frequencies and momenta.
With the $\chi^0_{{\bf G}{\bf G}'}$ matrices at hand, the interacting density-density response function~(susceptibility) $\chi_{{\bf G}{\bf G}'}$  was obtained from the central equation of TDDFT (see main text).
The latter is a Dyson-like equation, which can be rewritten as
\begin{equation}
\chi_{{\bf G}{\bf G}'}= \chi^0_{{\bf G}{\bf G}'}
+ \sum_{{\bf G}''{\bf G}'''}
\chi^0_{{\bf G}{\bf G}''}\,v_{{\bf G}''{\bf G}'''}\,\chi_{{\bf G}'''{\bf G}'},
\label{chieq}
\end{equation}
by making explicit the summation indices.

The effective potential matrix elements in Eq.~\eqref{chieq} may be calculated from the truncated Fourier integrals given in Eq.~\eqref{vTDC}, which yields
\begin{align}
\label{kernMat}
v_{{\bf G}{\bf G}'} =& \frac{4\pi\delta_{{\bf G}{\bf G}'}}{|{\bf q} + {\bf G}|^2}+
\frac{4\pi \delta _{{\bf G}{\bf G}'}
e^{-iL_{z}\frac{G-G'}{2}}}{L_{z}|\mathbf{q}+{\bf G}|} \\
&\qquad \quad \times \frac{1-e^{-L_{z}|\mathbf{q}+{\bf G}|}}{G^{2}+|\mathbf{q}+{\bf G}|^{2}} \,
\frac{GG'-|\mathbf{q}+{\bf G}|^{2}}{G'^{2}+|\mathbf{q}+{\bf G}|^{2}}.
\notag
\end{align}
The $v_{{\bf G}{\bf G}'}$ terms have been proved to efficiently cut off the spurious interaction between replicas of planar graphene-based systems~\cite{despoja2012ab,despoja2013two,despoja2015three,PRB_Pisarra}.

\begin{figure}[h]
\centerline{\includegraphics[width=0.47\textwidth]{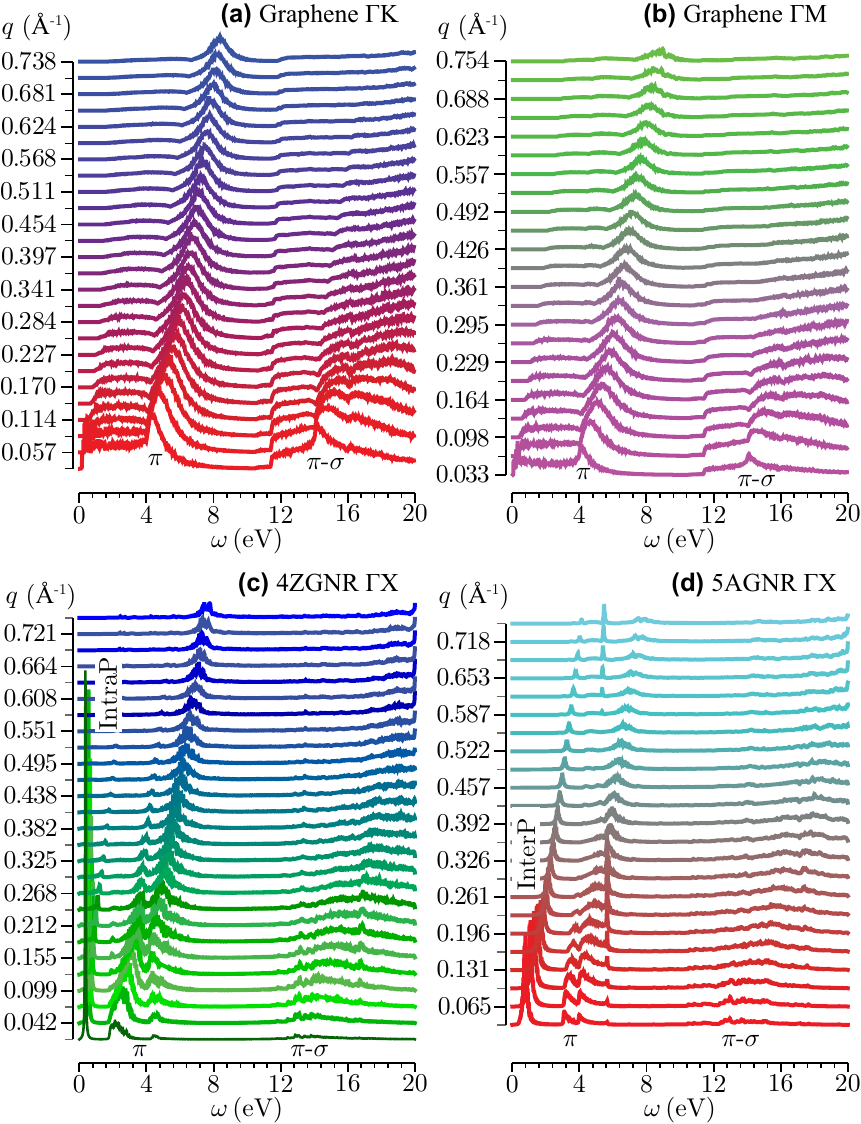}}
\vskip -10pt
\caption{
$E_{\msc{loss}}$ vs $\omega < 20$~eV for intrinsic MG [(a) and (b)], 4ZGNR (c) and 5AGNR (d) at $T=300$~K.
The EL spectra, corresponding to some of the sampled momenta in the density plots of Fig.~{\dplGNRsloss}, with $q \parallel \Gamma{K}$ (a), $q \parallel \Gamma{M}$ (b) and $q \parallel \Gamma{X}$ [(c) and (d)], are shifted vertically for clarity.
\label{FigS3}}
\end{figure}

\noindent Finally, the EL spectra were computed by
\begin{equation}
E_{\msc{loss}} = -\im\sum_{{\bf G}}v_{\mathbf{0} {\bf G}}\,\chi_{{\bf G} \mathbf{0}},
\end{equation}
where
\begin{equation}
\chi_{{\bf G} {\bf G}'}=\sum_{{\bf G}''}[(1-\chi^0 v)^{-1}]_{{\bf G} {\bf G}''}\chi^0_{{\bf G}''{\bf G}'},
\label{chisol}
\end{equation}
is the solution of Eq.~\eqref{chieq}.

Fig.~{\Intrinsic} provides a complementary view to Fig.~{\dplGNRsloss}, with the EL spectra of MG, 4ZGNR and 5AGNR being reported for selected $q$ values in the corresponding momentum domains.
As pointed out in the main text, the one-peak structure of the $\pi$ and $\sigma$-$\pi$ plasmons in MG is turned  into well resolved sub-peak structures in the GNRs.
These originate from the several  $\pi$ and $\sigma$ bands of the 1D materials~[Figs.~{\BSGnrs}(b), {\BSGnrs}(d), {\optgeom}(a) and {\optgeomB}(b)].

The low-energy intraband and interband modes for 4ZGNR and 5AGNR, respectively, appear to be well resolved and have higher intensities with respect to the $\pi$ and $\sigma$-$\pi$ modes.
In particular the intraband mode of 4ZGNR is generated by a conduction-electron concentration $n^*$ of $3.96{\times}10^{12}$~cm$^{-2}$~[Fig.~{\dplGNRsloss}(c)].

Such a mode is absent in MG at the absolute zero because of the vanishing behavior of the density of states~(DOS) at the inequivalent $K$ points of the first BZ.
The change in the Fermi-Dirac statistics at room temperature leads to  $n^*=1.15{\times}10^{11}$~cm$^{-2}$~[Figs.~{\dplGNRsloss}(a) and {\dplGNRsloss}(b)], which in turns is responsible for a weak structure in the optical region of MG. This is barely visible at the lowest $q$'s in the plots of Figs.~{\dplGNRsloss}(a), {\dplGNRsloss}(b), {\Intrinsic}(a) and {\Intrinsic}(b), due to both the on-set of the $\pi$ structure and the energy resolution~(0.01~eV) used to sample the $0-20$~eV range.
\begin{figure}[b]
\centerline{\includegraphics[width=0.47\textwidth]{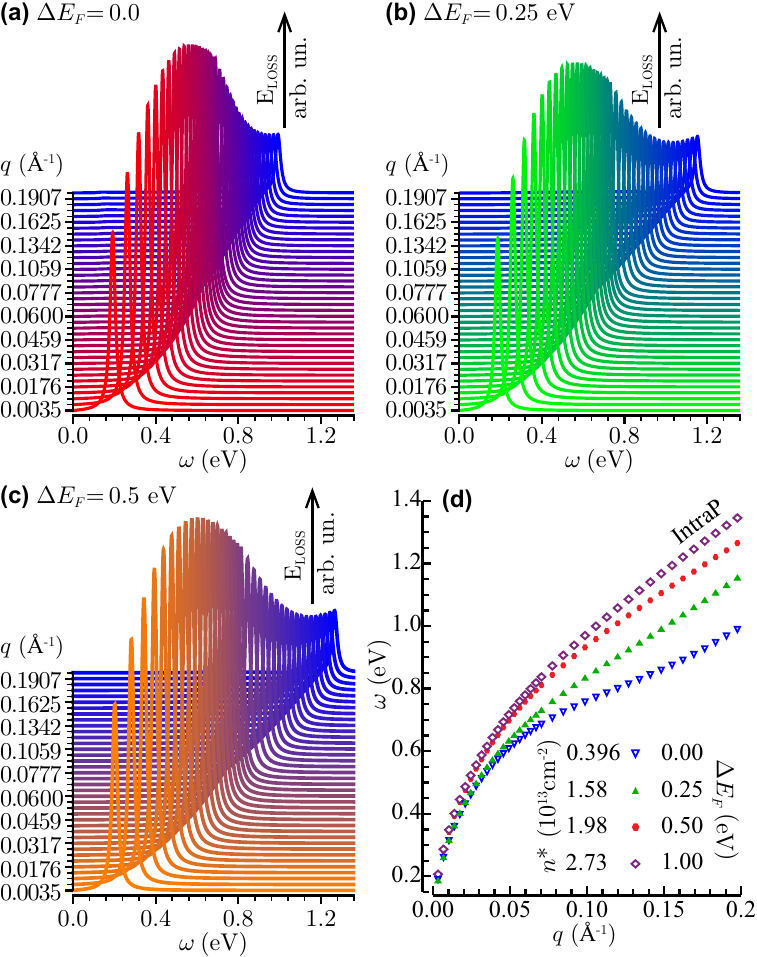}}
\vskip -10pt
\caption{
$E_{\msc{loss}}$ vs $\omega < 1.5$~eV (a)-(c) and intraband plasmon dispersion (d) for intrinsic and extrinsic 4ZGNR at $T=300$~K.
The EL spectra in (a)-(c), corresponding to some of the sampled momenta shown in the density plots of Figs.~{\DopedZGNR}(a)-{\DopedZGNR}(c), are shifted vertically for clarity. The plasmon resonances represented as green dots in Figs.~{\DopedZGNR}(a)-{\DopedZGNR}(d) are shown together in (d).
\label{FigS4}}
\end{figure}
\begin{figure*}
\centerline{\includegraphics[width=0.925\textwidth]{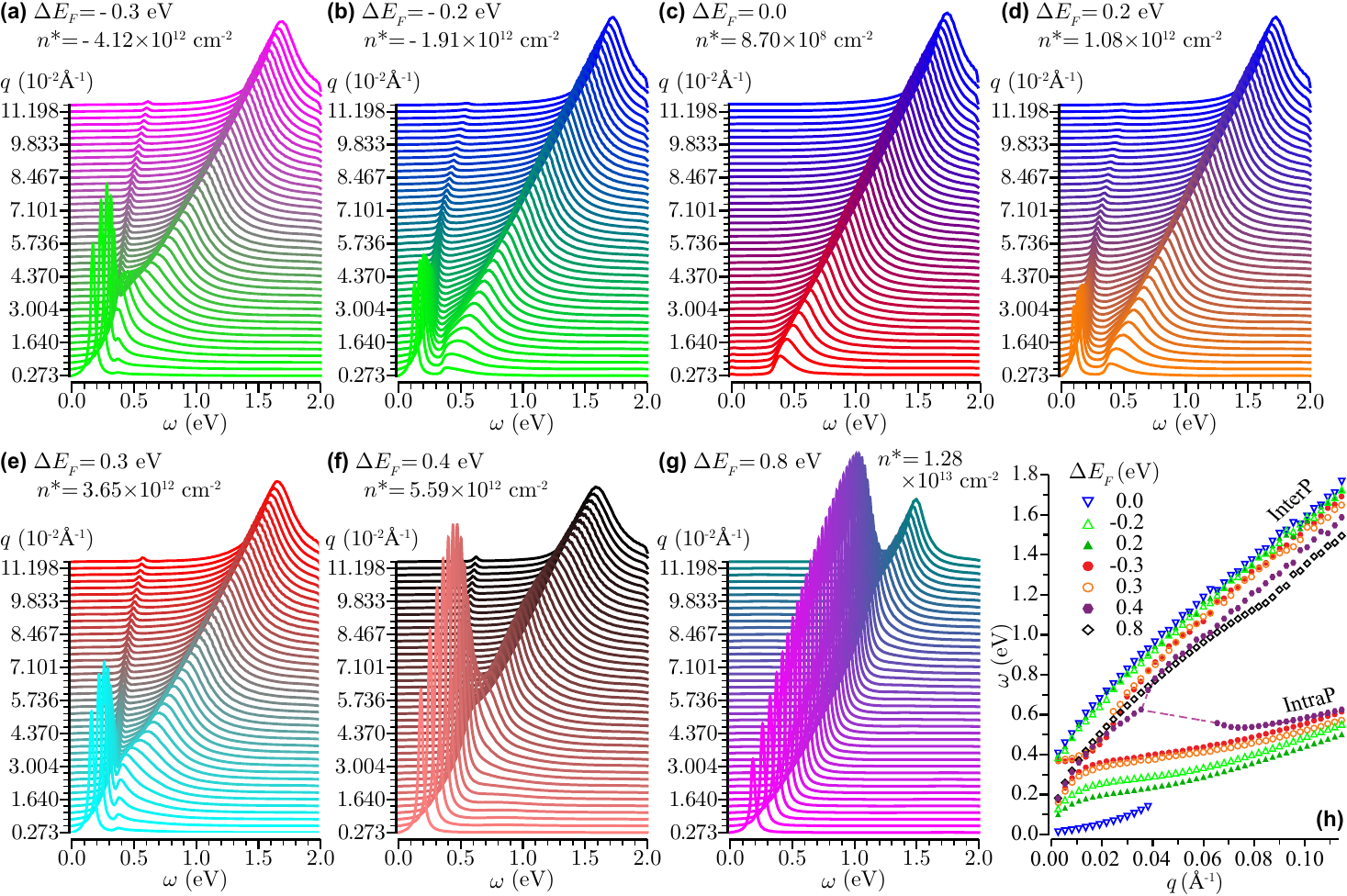}}
\vskip -10pt
\caption{
$E_{\msc{loss}}$ vs $\omega \leq 2$~eV (a)-(g) and $(q,\omega)$ plasmon dispersions (h) for intrinsic and extrinsic 5AGNR at $T=300$~K.
Values of $\Delta E_F$ ranging from $-0.3$ to $0.8$~eV are considered in (a)-(g), which correspond to charge-carrier concentrations $n^*$ of the order of $\pm 10^{12}$~cm$^{-2}$.
The EL spectra in (a)-(g) are represented for some of the sampled momenta below $0.15$~{\AA}$^{-1}$ along the GNR axis and shifted vertically for clarity.
\label{FigS5}}
\end{figure*}
\section{Doped Systems}
To explore the dielectric properties of the GNRs at infrared frequencies, we used an MP mesh of $1000\times1\times1$ wave vectors with $\sim$ 30 bands, covering the frequency region $\omega \leq 2$~eV.
Besides the dielectric properties of the pristine GNR arrays, we also studied how these properties change with doping.
Indeed doping can happen as an unwanted effect due to chemical contamination or it can be introduced in a controlled way, i.e., after functionalization of the GNR or through a gating potential.

Doping or gating  was simulated by changing the occupation factors $f_{\nu{\bf k}}$  and $f_{\nu{\bf k}+{\bf q}}$ in $\chi^0_{{\bf G}{\bf G}'}$~(see Eq.~\eqref{AdlWi}), according to the concentration of injected or ejected electrons.
It can be safely assumed that the doping values sampled in this Letter  have a negligible effect on the KS electronic structure~\cite{pisarra2014acoustic,PRB_Pisarra}.

All other settings were the same as for the intrinsic systems.
Figs.~{\DopedZGNRB} and~{\DopedAGNRB} provide a complementary representation of the EL properties of 4ZGNR, given in Figs.~{\DopedZGNR}(a)-{\DopedZGNR}(d), and 5AGNR, reported in Figs.~{\DopedAGNR}(a)-{\DopedAGNR}(d), respectively.

In semimetals like mono- and bilayer graphene, even a small doping yields the formation of a quasi two-dimensional gas of charge carriers (electrons or holes).
The fact that the conduction or valence band becomes partially occupied or unoccupied causes the appearance of low-energy intraband single-particle transitions and collective excitations.

In 4ZGNR, the non vanishing DOS at the Fermi level~[Fig.~{\BSGnrs}(b)] causes the appearance of a well resolved intraband mode being weakly affected by the doping (gating).
As shown in Fig.~{\DopedZGNRB}(d), the plasmon resonance energies slightly increase with increasing the injected electron concentration, following the typical squareroot-like dispersion of the two-dimensional plasmon in MG as $q\to0$.
These features are separately visible in Figs.~{\DopedZGNR}(a)-{\DopedZGNR}(d).

It is worth mentioning that the effective charge-carrier density contributing to the intraband mode in 4ZGNR is of the same order of magnitude for all sampled doping levels, i.e., $10^{13}$cm$^{-2}$. Therefore, moderate energy shifts, up to $0.5$~eV, are recorded in the intraband plasmon dispersions~[Fig.~{\DopedZGNRB}(d)].
In contrast, doping levels of $0.5$ and $1$~eV in MG lead to a more pronounced charge-carrier density increase, of $10^{13}$ and $10^{14}$ cm$^{-2}$, respectively.
Accordingly, a substantial variation in the plasmon energy of the extrinsic two-dimensional plasmon of MG has been  reported~\cite{pisarra2014acoustic}.

In 5AGNR, the presence of a small gap with high DOS~[Fig.~{\BSGnrs}(d)] allows the appearance of an interband plasmon that seems to be weakly influenced by electron injection or ejection.

On the other hand, positive or negative Fermi energy shifts generate an intraband plasmon that is extremely sensitive to the doping level and may interfere with the interband plasmon.
These other features are highlighted in Fig.~{\DopedAGNRB}(h), where the intraband and interband plasmon dispersions of Figs.~{\DopedAGNRB}(a)-{\DopedAGNRB}(g) are plotted together vs $q$ and $\omega$.
Some of them are also separately visible in Figs.~{\DopedAGNR}(a)-{\DopedAGNR}(d).
Notice that a tiny intraband contribution should be present in undoped 5AGNR because of the Fermi-Dirac population of low lying conduction states at room temperature.

Indeed, 5ANGR is characterized by a conduction-electron concentration of ${\sim}10^9$~cm$^{-2}$ at $T=300$~K, which is negligibly small with respect to that of 4ZGNR and MG. Nonetheless, local maxima in $E_{\msc{loss}}$ are recorded for $q < 0.04$~\AA$^{-1}$, yielding the $(\omega,{\bf q})$ dispersion in Fig.~{\DopedAGNRB}(h).
Also detectable in Fig.~{\DopedAGNRB}(h) is the effect of the slight asymmetry in the valence and conduction bands of 5AGNR close to band gap~[inset in Fig.~{\BSGnrs}(d)], which appears to influence the plasmon structure especially in the intra-band components. This is due to the different charge-carrier concentrations originated from positive and negative doping values.

Thus, the intraband plasmon of 5AGNR depends of the doping type~(sign) even at frequencies of the order of $\sim 50$~THz.

We should also mention that in a realistic situation of suspended GNR arrays on top of a substrate, Fermi level shifts of $0.5-1$~eV, obtained for example by applying a gate voltage in air, can lead to a collapsing of the sample.
Nevertheless, the carrier density concentration predicted in our 4ZGNR--with a doping level of $1.0$~eV--is similar to the experimental concentration of $2.5{\times}10^{13}$~cm$^{-2}$ achieved  by depositing alternating wafer-scale graphene sheets and thin insulating layers~\cite{ib03}. Furthermore, electrical conductance measurements at small and room temperatures, performed on a set of doped parallel GNRs, with widths ranging from 14 to 63 nm, has given a charge-carrier concentration of $3.6{\times}10^{12}$~cm$^{-2}$~\cite{ib02}. This value is in close agreement with the predicted charge carrier concentration of our doped 5AGNR arrays~(Figs.~{\DopedAGNR} and~{\DopedAGNRB}).
\begin{figure}[!h]
\centerline{\includegraphics[width=0.47\textwidth]{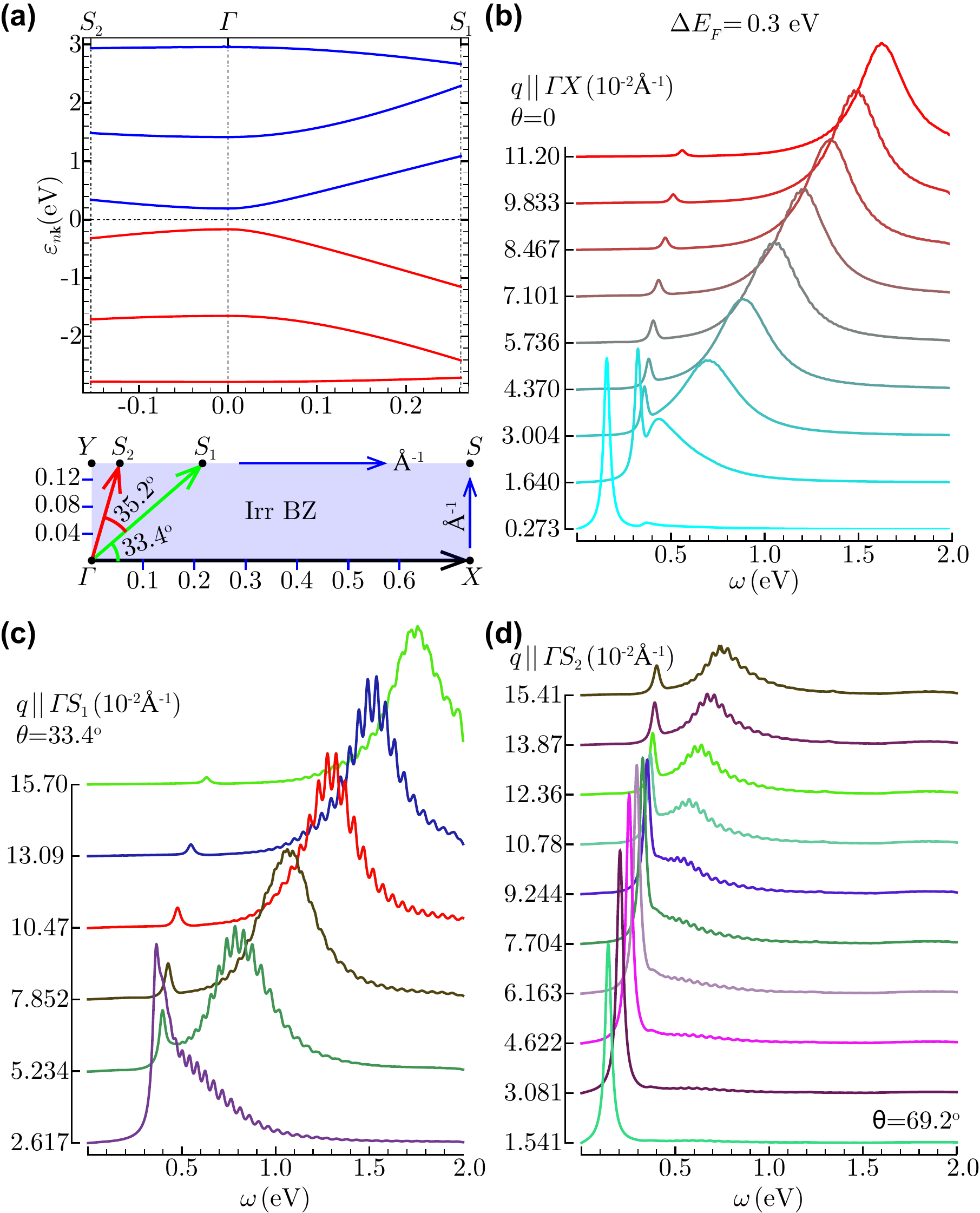}}
\vskip -20pt
\caption{
Band structure (a, Top), first BZ (a, Bottom) and loss spectrum of extrinsic 5AGNRs at $T=300$~K, $\Delta E_F=0.3$~eV. The $E_{\msc{loss}}$ curves are obtained by re-sampling first BZ with an MP mesh of $270{\times}40{\times}1$ points. These are shown vs $\omega < 2$~eV for some incident momenta along different directions, forming angles of $\theta=0$~($\Gamma{X}$, b), $33.4$ ($\Gamma{S}_1$, c) and $69.2^{\circ}$ ($\Gamma{S}_2$, d) relative to the ribbon axis.
\label{FigS6}}
\vskip -10pt
\end{figure}

The interplay between the intraband and interband modes can be also detected by probing 5ANGR with oblique incident momenta of in-plane modulus $q$ forming different angles $\theta$ relative to the ribbon axis.
This is shown in Fig.~{\DopedAGNROB} where the EL spectra have been computed by re-sampling the first BZ of the system with an MP mesh of $270{\times}40{\times}1$ points.
We see that specific combinations of modulus and angle in the incident momentum lead the two modes to be superimposed, whereas other choices leave the two-peak structure separated in momentum space.
\begin{figure*}
\centerline{\includegraphics[width=0.925\textwidth]{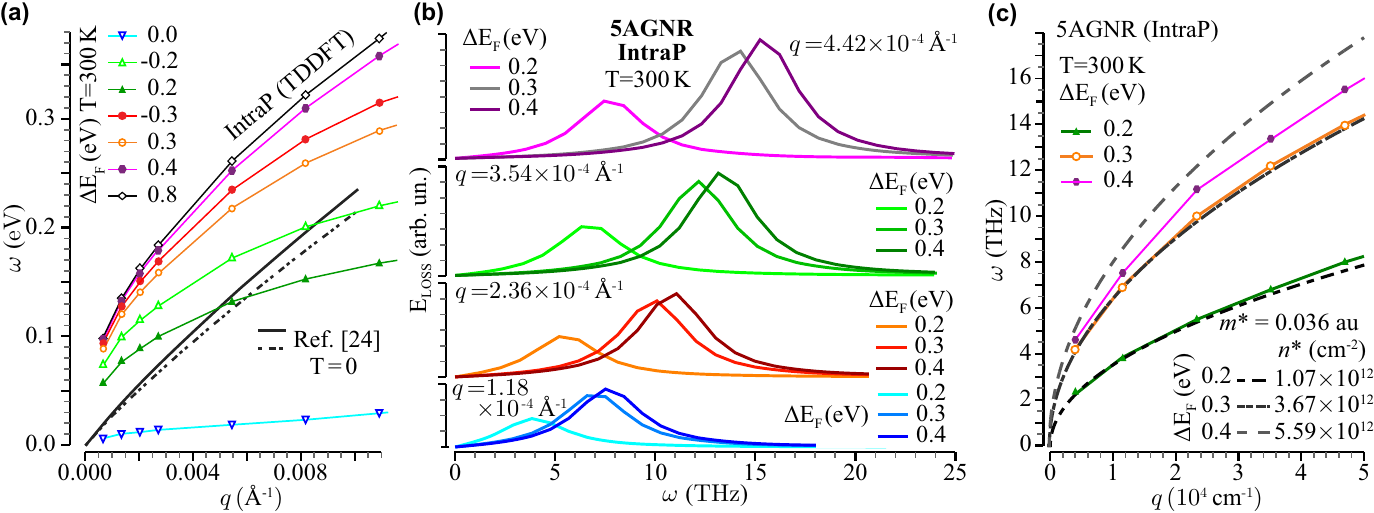}}
\vskip -10pt
\caption{
Intraband plasmon dispersions [(a) and (c)] and EL spectra (b) for 5AGNR at room temperature, low frequencies~($\omega < 100$~THz), and small momenta~($q <10^5$~cm$^{-1}$).
TDDFT calculations with the same settings as Fig.~{\DopedAGNRB} are shown in (a) and compared with the TB calculations of Ref.~\cite{andersen2012plasmon}~(black continuous and dashed lines),  where two different localization parameters where chosen to model the electron-electron interaction entering Eq.~\eqref{chieq}.
The EL spectra in (b) are computed with a TDDFT approach that include only the valence and conduction states.
The intraband plasmon dispersions of (b) are shown in (c) and compared with the semiclassical model of Ref.~\cite{popov2010oblique} (dashed lines)
\label{FigS7}}
\end{figure*}

The role of positive doping is also spotted in Fig.~{\THzAGNRB}(a), where a comparison of the intraband plasmon dispersion of 5AGNR is attempted with the calculations of Ref.~\cite{andersen2012plasmon}.
The latter were based on a tight-binding~(TB) gapless two-band model at the absolute zero. A two-dimensional Coulomb potential was adopted, depending on a localization parameter \Ignore{TBpz-w} of the $p_z$ orbitals.
The correlation matrix elements~\eqref{rhocc} were approximated by the low-$q$ limit form $\rho_{\nu\nu'}^{k{\bf q}}({\bf 0}) \approx \braket{\nu {\bf k}}{\nu' {\bf k} + {\bf q}}$. Of course, the detailed properties of the plasmon modes are critically dependent on the valence and conduction band dispersions, which are significantly different in DFT-LDA and TB calculations.
The TB electronic features of 5AGNR allow only for intraband modes that, nevertheless, appear at the same scale as the TDDFT intraband plasmons of doped 5AGNR.

To have a closer look at the THz region, we performed TDDFT calculations on an MP mesh of $12500\times1\times1$ points, including only the contributions of valence and conduction states of 5AGNR in Eq.~\eqref{AdlWi}. Well converged results were obtained with a broadening lifetime parameter $\eta$ of $\sim 0.5$~THz.

The EL spectra for 5AGNR at $T=300$~K and $\omega < 25$~THz are shown in Fig.~{\THzAGNRB}(b).
The plasmon dispersion curves are reported in Fig.~{\THzAGNRB}(c) and compared to the semiclassical approach of Ref.~\cite{popov2010oblique}, where plasmon resonances in doped GNR arrays of $10-100$ nm width were described by the following analytical expression:
\begin{equation}
\omega_{p}=\re(\sqrt{2 \pi n^* q \cos ^2\theta/m^*-\eta ^2}-i\eta)
\label{omeP}
\end{equation}
Eq.~\eqref{omeP}~(given in Hartree au) depends on the concentration  $n^*$ of conduction electrons, whose effective mass $m^*$ is directly proportional to the band gap $E_{\msc{gap}}$ through the characteristic velocity $v_{0} \sim 10^6$~m/s, i.e., $E_{\msc{gap}}=2 m^* v_{0}$.
To use Eq.~\eqref{omeP}, we estimated $m^*\sim0.36$ corresponding to $E_{\msc{gap}}\sim0.4$~eV and considered  concentration values equivalent to the Fermi energy shifts $\Delta E_{F}=0.2, 0.3, 0.4$~eV at $T=300$~K.
The analytical and numerical curves turned out to be amazingly similar for $\Delta E_{F} < 0.5$~eV.

Then, our TDDFT calculations on a narrow-width GNR have the correct $q{\to}0$,$\omega{\to}0$-limit that fits with (non-{\it ab initio}) predictions on larger GNR structures currently available for experiments.
Such TB approaches, due to their simplified description of the band energies and wave functions, are only able to predict the existence of one plasmon mode, the extrinsic intraband plasmon.

\section{Temperature effects in 5AGNR}

As shown in the main text and in the previous section, the large tunability in the intraband plasmon mode of 5AGNR is critically dependent on the electron or hole occupancy of  conduction or valence states, lying within an energy window of $0.5-1$~eV around the Fermi level.
Such a population is given by the Fermi-Dirac statistical factors $f_{\nu{\bf k}}$ and $f_{\nu{\bf k}+{\bf q}}$ in Eq.~\eqref{AdlWi} that are significantly influenced, within the considered energy range, by even moderate temperature changes below $\sim 1000$~K.

This effect inevitably plays a major role in any nanodevice design approach.
To quantify and characterize it, we ran EL calculations on undoped 5AGNR with electronic temperature values larger than $500$~K.
All other settings were the same as for room temperature calculations.
The resulting EL spectra are reported in Fig.~{\AGNRtemp} for energies below $\sim 1$~eV and momenta smaller than $0.03$~${\AA}^{-1}$.

At $T=300$ K~[Fig.~{\AGNRtemp}(a)], the interband plasmon is clearly visible. Nevertheless a faint intraband peak may be spotted at energies below $\sim 0.1$~eV, and the intraband plasmon dispersion can be computed.
As the temperature is increased to $500$~K~[Fig.~{\AGNRtemp}(b)], the intraband plasmon peak begins to appear in the same intensity scale as the interband plasmon.
At higher temperatures, say, $T=700,900$ K~[Figs.~{\AGNRtemp}(c) and {\AGNRtemp}(d)], the intraband plasmon is well resolved and also well separated from the interband plasmon.

The increase of the intraband plasmon intensity  is readily understood considering how the electronic temperature affects the population of the KS states near the Fermi level.
At room temperature the number of electrons that are capable to overcome the $0.36$~eV gap is small, with a concentration of the order of $10^9$ cm$^{-2}$. Thus, interband excitations are dominant.
At $T=500$ K the electron population of the conduction band becomes appreciable, with a concentration of  roughly $2{\times}10^{10}$ cm$^{-2}$, generating the small peak in the EL spectrum.
As the temperature further increases~($T\geq700$ K), the smearing width of the Fermi-Dirac distribution function increases, the conduction electron concentrations become larger than $10^{11}$ cm$^{-2}$, and the intraband plasmon fully appears in the EL spectrum.
Charge-carrier concentrations triggered by temperature increase are nevertheless much smaller than those obtained with doping or gating. For this reason, no particular interference is recorded in Fig.~{\AGNRtemp} between intraband and interband plasmon modes. It should be also mentioned that in the realistic case of GNR arrays suspended on top of a substrate, the substrate phonons become relevant and significantly affect the plasmon damping at temperatures above $500$~K.

\begin{figure*}
\centerline{\includegraphics[width=0.925\textwidth]{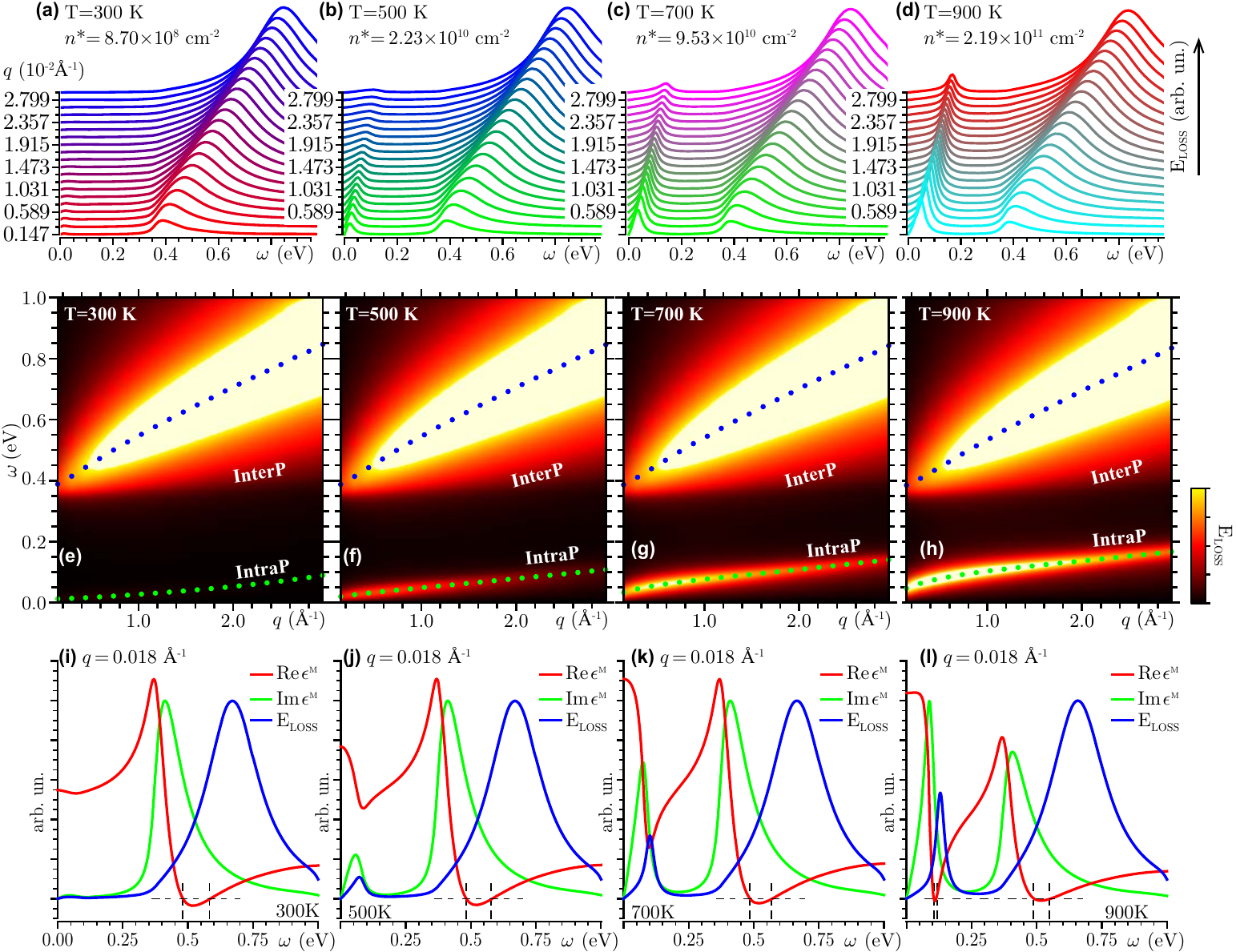}}
\vskip -10pt
\caption{
Dielectric response of intrinsic 5AGNR at $T=300$ K [(a),(e) and (i)],  $T=500$ K [(b), (f) and (j)], $T=700$ K [(c),(g) and (k)], $T=900$ K [(d), (h) and (l)].
$E_{\msc{loss}}$ is represented as sequence of shifted spectra in (a)-(d) and density plots in (e)-(h) with the same intensity scale as in Fig.~{\dplGNRsloss}, Figs.~{\DopedZGNR}(a)-{\DopedZGNR}(d) and Figs.~{\DopedAGNR}(a)-{\DopedAGNR}(d). The green and blue dots denote the intraband and interband plasmon $(q,\omega)$-dispersions, respectively.
$\re(\epsilon^{\msc{m}})$, $\im(\epsilon^{\msc{m}})$ and $E_{\msc{loss}}$ in (i)-(l) are normalized as in Figs.~{\DopedZGNR}(e)-{\DopedZGNR}(h) and Figs~{\DopedAGNR}(e)-{\DopedAGNR}(h).
The dashed grid-lines indicate the zeros of $\re(\epsilon^{\msc{m}})$.
\label{FigS8}}
\end{figure*}
\section{Final remarks}
We have discussed the response of planar GNR arrays with armchair and zigzag shaped edges to a probe particle~(i.e, an electron or a photon) of energy below $20$~eV and in-plane momentum smaller than $0.8~$\AA $^{-1}$.
The use of an {\it ab initio} strategy based on TDDFT in the RPA has let us characterize both the intrinsic and extrinsic plasmon structures of the systems, which could not have possibly been predicted by less sophisticated TB approaches.
At high-energies~($> 3$~eV), we have found the two standard interband excitations of carbon-based materials, namely the $\pi$ and $\sigma$-$\pi$ plasmons, whose energy-momentum dispersions are strongly influenced by the GNR geometry.
At lower energies~($< 1.5$~eV), we have detected one intraband plasmon mode for the semimetallic ZGNR.
In the same energy range, we have identified two modes in the semiconducting AGNR, i.e., one interband and  one intraband plasmons.
The latter shows a large sensitivity to doping level as well as electronic temperature.
The interplay between the two collective excitation in AGNR, which co-exist and interact with each other, represent the main result of our study.
A similar interference effect, between a two-dimensional plasmon and acoustic-like plasmon, has been reported in a recent study on bilayer graphene~\cite{ibxx}.
We have checked that the intraband and interband plasmons appear also in larger AGNRs.
These findings require further confirmation by measurements, although the experiments of Ref.~\cite{Feinanolett.5b03834} seem to suggest that interband and intraband plasmon in low-gap GNRs exist, can be controlled and exploited for future nanodevice technology.

%
%
%
%

\clearpage

\begin{thebibliography}{99}
\vskip 12pt
\textbf{References}
\bibitem{Lin2009}
Y.-M. Lin, K.~A. Jenkins, A.~Valdes-Garcia, J.~P. Small, D.~B. Farmer and P.~Avouris,
\newblock Nano Lett. {\bf 9}, 422 (2009).

\bibitem{Shuba2009}
M.~V. Shuba, G.~Y. Slepyan, S.~A. Maksimenko, C.~Thomsen and A.~Lakhtakia,
\newblock Phys. Rev. B {\bf 79}, 155403 (2009).

\bibitem{bao2012graphene}
Q.~Bao and K.~P. Loh,
\newblock ACS Nano {\bf 6}, 3677 (2012).

\bibitem{sensale2012graphene}
B.~Sensale-Rodriguez, R.~Yan, M.~M. Kelly, T.~Fang, K.~Tahy, W.~S. Hwang, D.~Jena, L.~Liu and H.~G. Xing,
\newblock Nat. Commun. {\bf 3}, 780 (2012).

\bibitem{ThongrattanasiriACSNANO2012}
S.~Thongrattanasiri, A.~Manjavacas and F.~J.~G. de~Abajo,
\newblock ACS Nano {\bf 6}, 1766 (2012).

\InT{
\bibitem{ib00}
Yu. V. Bludov, A. Ferreira, N. M. R. Peres and M. I. Vasilevskiy,
\newblock Int. J. Mod. Phys. {\bf B27}, 1341001 (2013).

\bibitem{ib01}
T. Low and P. Avouris,
\newblock ACS Nano {\bf 8}(2), 1086 (2014).
}

\bibitem{garcia2014graphene}
F.~J. Garcia~de Abajo,
\newblock ACS Photonics {\bf 1}, 135 (2014).

\bibitem{brun2015}
C.~Brun, T.~C. Wei, P.~Franck, Y.~C. Chong, L.~Congxiang, C.~W. Leong, D.~Tan,
  T.~B. Kang, P.~Coquet and D.~Baillargeat,
\newblock IEEE Trans. THz Sci. Technol. {\bf 5}, 383 (2015).

\bibitem{fei2015} Z. Fei, E. G. Iwinski†, G. X. Ni, L. M. Zhang, W. Bao, A. S. Rodin, Y. Lee, M. Wagner, M. K. Liu, S. Dai, M. D. Goldflam, M. Thiemens, F. Keilmann, C. N. Lau, A. H. Castro-Neto, M. M. Fogler, and D. N. Basov,
\newblock Nano Lett. {\bf 15}(8), 4973 (2015).

\bibitem{neto2009electronic}
A.~C. Neto, F.~Guinea, N.~Peres, K.~S. Novoselov and A.~K. Geim,
\newblock Rev. Mod. Phys. {\bf 81}, 109 (2009).

\bibitem{ju2011graphene}
L.~Ju {\em et~al.},
\newblock Nat. Nanotechnol.  {\bf 6}, 630 (2011).

\bibitem{zhou2012atomically}
W.~Zhou, J.~Lee, J.~Nanda, S.~T. Pantelides, S.~J. Pennycook and J.-C. Idrobo,
\newblock Nat. Nanotechnol.  {\bf 7}, 161 (2012).

\bibitem{yan2013damping}
H.~Yan, T.~Low, W.~Zhu, Y.~Wu, M.~Freitag, X.~Li, F.~Guinea, P.~Avouris and
  F.~Xia,
\newblock Nat. Photonics {\bf 7}, 394 (2013).

\bibitem{liou2015pi}
S.~C. Liou, C.-S. Shie, C.~H. Chen, R.~Breitwieser, W.~W. Pai, G.~Y. Guo and
  M.-W. Chu,
\newblock Phys. Rev.  B {\bf 91}, 045418 (2015).

\bibitem{cupolillo2015substrate}
A.~Cupolillo, A.~Politano, N.~Ligato, D.~C. Perez, G.~Chiarello and L.~Caputi,
\newblock Surface Science {\bf 634}, 76 (2015).

\bibitem{pisarra2014acoustic}
M.~Pisarra, A.~Sindona, P.~Riccardi, V.~M. Silkin and J.~M. Pitarke,
\newblock New Journal of Physics {\bf 16}, 083003 (2014).

\bibitem{PRB_Pisarra}
M.~Pisarra, A.~Sindona, M.~Gravina, V.~M. Silkin and J.~M. Pitarke,
\newblock Phys. Rev. B {\bf 93}, 035440 (2016).

\bibitem{Sind_SPRING} A. Sindona, M. Pisarra, D. Mencarelli, L. Pierantoni, S. Bellucci, ``Plasmon Modes in Extrinsic Graphene: Ab initio Simulations vs Semi-classical Models", pp.125-144, Chap. 7, ``Fundamental and Applied Nano-Electromagnetics'',  Springer 2016 (A. Maffucci, S. A. Maksimenko, ed.)

\InT{\bibitem{ibxx} T. Low, F. Guinea, H. Yan, F. Xia, and P. Avouris,
\newblock Phys. Rev. Lett. 112, 116801 (2014)}

\bibitem{christensen2011graphene}
J.~Christensen, A.~Manjavacas, S.~Thongrattanasiri, F.~H. Koppens and F.~J.
  Garc{\'{i}}a~de Abajo,
\newblock ACS Nano {\bf 6}, 431 (2011).

\bibitem{GraBN_plasmon}
A.~Woessner {\em et~al.},
\newblock Nature Materials {\bf 14}, 421–425 (2015).

\bibitem{Tong2015}
J.~Tong, M.~Muthee, S.-Y. Chen, S.~K. Yngvesson and J.~Yan,
\newblock Nano Lett. {\bf 15}, 5295 (2015).

\bibitem{nikitin2012surface}
A.~Y. Nikitin, F.~Guinea, F.~J. Garcia-Vidal and L.~Martin-Moreno,
\newblock Phys. Rev. B {\bf 85}, 081405 (2012).

\bibitem{Feinanolett.5b03834}
Z.~Fei {\em et~al.},
\newblock Nano Lett. {\bf 15}, 8271 (2015).

\bibitem{son2006energy}
Y.-W. Son, M.~L. Cohen and S.~G. Louie,
\newblock Phys. Rev. Lett. {\bf 97}, 216803 (2006).

\bibitem{zGNRGAP2007}
L.~Yang, C.-H. Park, Y.-W. Son, M.~L. Cohen and S.~G. Louie,
\newblock Phys. Rev. Lett. {\bf 99}, 186801 (2007).

\bibitem{Dubois2009}
S.~M.~M. Dubois, Z.~Zanolli, X.~Declerck and J.~C. Charlier,
\newblock Eur. Phys. J. B {\bf 72}, 1 (2009).

\bibitem{kan2008half}
E.-j. Kan, Z.~Li, J.~Yang and J.~Hou,
\newblock Journal of the American Chemical Society {\bf 130}, 4224 (2008).

\bibitem{tao2011spatially}
C.~Tao {\em et~al.},
\newblock Nature Physics {\bf 7}, 616 (2011).

\bibitem{popov2010oblique}
V.~V. Popov, T.~Y. Bagaeva, T.~Otsuji and V.~Ryzhii,
\newblock Phys. Rev. B {\bf 81}, 073404 (2010).

\bibitem{BreyETAl2006}
L.~Brey and H.~A. Fertig,
\newblock Phys. Rev. B {\bf 73}, 235411 (2006).

\bibitem{andersen2012plasmon}
D.~R. Andersen and H.~Raza,
\newblock Phys. Rev. B {\bf 85}, 075425 (2012).

\bibitem{supI}
See Supplemental Material for the following:
DFT calculations, geometry optimizations (Sec. I);
TDDFT scheme, intrinsic GNRs (Sec. II);
doped GNRs, interband and intraband plasmon dispersions (Sec. III);
temperature effects (Sec. IV).

\bibitem{perdew1981self}
J.~P. Perdew and A.~Zunger,
\newblock Phys. Rev. B {\bf 23}, 5048 (1981).

\bibitem{troullier1991efficient}
N.~Troullier and J.~L. Martins,
\newblock Phys. Rev. B {\bf 43}, 1993 (1991).

\bibitem{Abinit}
X.~Gonze {\em et~al.},
\newblock Comput. Phys. Commun. {\bf 180}, 2582 (2009).

\bibitem{monkhorst1976special}
H.~J. Monkhorst and J.~D. Pack,
\newblock Phys. Rev. B {\bf 13}, 5188 (1976).

\bibitem{adler1962quantum}
S.~L. Adler,
\newblock Phys. Rev. {\bf 126}, 413 (1962).

\bibitem{wiser1963dielectric}
N.~Wiser,
\newblock Phys. Rev. {\bf 129}, 62 (1963).

\bibitem{kubo1957statistical}
R.~Kubo,
\newblock J. Phys. Soc. Jpn.  {\bf 12}, 570 (1957).

\bibitem{PhysRevB.68.205106}
V.~M. Silkin, E.~V. Chulkov and P.~M. Echenique,
\newblock Phys. Rev. B {\bf 68}, 205106 (2003).

\bibitem{TDDFT}
M.~Petersilka, U.~J. Gossmann and E.~K.~U. Gross,
\newblock Phys. Rev. Lett. {\bf 76}, 1212 (1996).

\bibitem{onida2002electronic}
G.~Onida, L.~Reining and A.~Rubio,
\newblock Rev. Mod. Phys. {\bf 74}, 601 (2002).

\bibitem{despoja2012ab}
V.~Despoja, K.~Dekani{\'c}, M.~{\v{S}}unji{\'c} and L.~Maru{\v{s}}i{\'c},
\newblock Phys. Rev. B {\bf 86}, 165419 (2012).

\bibitem{despoja2013two}
V.~Despoja, D.~Novko, K.~Dekani{\'c}, M.~{\v{S}}unji{\'c} and
  L.~Maru{\v{s}}i{\'c},
\newblock Phys. Rev. B {\bf 87}, 075447 (2013).

\bibitem{despoja2015three}
D.~Novko, V.~Despoja and M.~{\v{S}}unji{\'c},
\newblock Phys. Rev. B {\bf 91}, 195407 (2015).


\bibitem{PhysRevLett.100.196803}
C. Kramberger, R. Hambach, C. Giorgetti, M. H. R\"{u}mmeli, M. Knupfer, J. Fink, B. B\"{u}chner, Lucia Reining, E. Einarsson, S. Maruyama, F. Sottile, K. Hannewald, V. Olevano, A. G. Marinopoulos, and T. Pichler,
\newblock Phys. Rev. Lett. {\bf 100}, 196803 (2008).

\bibitem{PhysRevB.77.233406}
T.~Eberlein, U.~Bangert, R.~R. Nair, R.~Jones, M.~Gass, A.~L. Bleloch, K.~S.
  Novoselov, A.~Geim and P.~R. Briddon,
\newblock Phys. Rev. B {\bf 77}, 233406 (2008).

\bibitem{PhysRevB.88.075433}
P.~Wachsmuth, R.~Hambach, M.~K. Kinyanjui, M.~Guzzo, G.~Benner and U.~Kaiser,
\newblock Phys. Rev. B {\bf 88}, 075433 (2013).

\bibitem{PhysRevB.90.235434}
P.~Wachsmuth, R.~Hambach, G.~Benner and U.~Kaiser,
\newblock Phys. Rev. B {\bf 90}, 235434 (2014).

\bibitem{taft}
E.~A. Taft and H.~R. Philipp,
\newblock Phys. Rev. {\bf 138}, A197 (1965).



\bibitem{RevModPhys.54.437}
T. Ando, A. B. Fowler, and F. Stern,
\newblock Rev. Mod. Phys. {\bf 54}, 437 (1982).

\InT{
\bibitem{ib03}
H. Yan, X. Li, B. Chandra, G. Tulevski, Y. Wu, M. Freitag, W. Zhu, P. Avouris and F. Xia,
\newblock Nat. Nanotechnol.  {\bf 7}, 330 (2012)

\bibitem{ib02}
M. Y. Han, B. \"{O}zyilmaz, Y. Zhang, and P. Kim,
\newblock Phys. Rev. Lett. {\bf 98}, 206805 (2007).
}
\end{thebibliography}
\end{document}